\documentclass[a4paper]{article}

\usepackage[english]{babel}
\usepackage[utf8x]{inputenc}
\usepackage[T1]{fontenc}

\usepackage[a4paper,top=3cm,bottom=2cm,left=3cm,right=3cm,marginparwidth=1.75cm]{geometry}

\usepackage{amsmath}
\usepackage{algpseudocode}
\usepackage{graphicx}
\usepackage{epsfig}
\usepackage{float}
\usepackage{makecell}
\usepackage{subcaption}
\usepackage[colorinlistoftodos]{todonotes}
\usepackage[colorlinks=true, allcolors=blue]{hyperref}

\title{CoHSI V: Identical multiple scale-independent systems within genomes and computer software}
\author{Les Hatton\footnote{Emeritus Professor, Kingston University, KT1 2EE, U.K., lesh@oakcomp.co.uk}, Gregory Warr\footnote{Emeritus Professor, Medical University of South Carolina, 171 Ashley Ave, Charleston, SC 29425, USA, gwawarr@gmail.com}}

\begin{document}

\maketitle

\begin{abstract}
A mechanism-free and symbol-agnostic conservation principle, the Conservation of Hartley-Shannon Information (CoHSI) is predicted to constrain the structure of discrete systems regardless of their origin or function. Despite their distinct provenance, genomes and computer software share a simple structural property; they are linear symbol-based discrete systems, and thus they present an opportunity to test in a comparative context the predictions of CoHSI. Here, \textit{without any consideration of, or relevance to, their role in specifying function} we identify that 10 representative genomes (from microbes to human) and a large collection of software contain identically structured nested subsystems. In the case of base sequences in genomes, CoHSI predicts that if we split the genome into n-tuples (a 2-tuple is a pair of consecutive bases; a 3-tuple is a trio and so on), \textit{without regard for whether or not a region is coding}, then each collection of n-tuples will constitute a homogeneous discrete system and will obey a power-law in frequency of occurrence of the n-tuples.  We consider 1-, 2-, 3-, 4-, 5-, 6-, 7- and 8-tuples of ten species and demonstrate that the predicted power-law behavior is emphatically present, and furthermore as predicted, is insensitive to the start window for the tuple extraction i.e. the reading frame is irrelevant.

We go on to provide a proof of Chargaff's second parity rule and on the basis of this proof, predict higher order tuple parity rules which we then identify in the genome data.

CoHSI predicts precisely the same behavior in computer software.  This prediction was tested and confirmed using 2-, 3- and 4-tuples of the hexadecimal representation of machine code in multiple computer programs, underlining the fundamental role played by CoHSI in defining the landscape in which discrete symbol-based systems must operate.
\end{abstract}

\section*{Statement of reproducibility}
This paper adheres to the transparency and reproducibility principles espoused by \cite{Popper1959,Ziolkowski1982,Claerbout1992,HatRob94,Donoho2009,Ince2012} and includes references to all methods and source code necessary to reproduce the results presented.  These are referred to here as the \textit{reproducibility deliverables} and will be available initially at https://leshatton.org.  Each reproducibility deliverable allows all results, tables and diagrams to be reproduced individually for that paper, as well as performing verification checks on machine environment, availability of essential open source packages, quality of arithmetic and regression testing of the outputs \cite{HattonWarr2016}.  Note that these packages are designed to run on Linux machines for no other reason than to guarantee the absence of any closed source and therefore potentially opaque contributions to these results.
\section*{Introduction}
In considering both genomes and computer software  we emphasize that the properties we examine in this study have nothing to do with the function of these systems in conveying information respectively to living organisms and to computers. Although interesting, the analogy of cells as living computers \cite{Brenner2012,Witzany2012,Condon2018}  is therefore without any relevance to the properties we consider here. The conservation principle (CoHSI, Conservation of Hartley-Shannon Information) that we have described is applicable to \textit{all discrete systems}; CoHSI accurately predicts amongst other \textit{global} properties, the length distribution of components in all discrete systems, including proteomes and other qualifying discrete systems such as software, written texts and musical compositions, at all levels of aggregation \cite{HattonWarr2015,HattonWarr2017,HattonWarr2018a,HattonWarr2018d}. Thus, of necessity in describing systems of such diverse provenance, CoHSI is symbol agnostic (in our usage "symbol" is synonymous with "sign" or "token") and the use of Hartley-Shannon Information, which eschews any meaning associated with  symbols, is integral to the theory.   In proteins, the symbols  are the amino acids (including any post-translational modifications) and in software they are the human-readable programming language tokens familiar to computer programmers.   As noted above, CoHSI is agnostic with respect to symbol but requires that the categorization of symbols within a system should be consistent.  In other words, we should see CoHSI-constrained properties regardless of the type or level of categorization used.  This we can examine in both proteins and in computer software because there is a deeper and very well known layer of categorization in both these systems; we refer respectively to the bases of DNA and to machine-readable code.

Proteins are encoded in DNA by (A)denine, (T)hiamine, (G)uanine and (C)ytosine, whereby, read sequentially, each triplet of bases encodes an amino acid.  Genomes consists of linear polymers of the bases and in complex organisms not all DNA is coding. Whereas in bacteria over 90 percent of the genome can encode proteins, in  humans for example the coding regions are typically considered to be between 1-2 percent of the total, and the function of the remaining DNA, which includes substantial amounts of repetitive sequences as well as regions regulating gene expression, has stirred controversy \cite{Doolittle5294}.  Note that we do not consider here the non-canonical bases, for example, the modified guanine 7-methylguanine.  The 17 known such modifications at the time of writing\footnote{https://en.wikipedia.org/wiki/DNA, accessed 17-Feb-2019.} would of course increase the base alphabet significantly from just 4, but sufficient data is not yet available for a comprehensive analysis incorporating this extended base alphabet.  For the record, since our theory is independent of the nature of the bases, we would not anticipate any changes to our conclusions as the base alphabet was extended.

Similarly, computer software systems are written in a higher-level human-readable form in one or more programming languages such as C, C++, perl or python.  However, before it can actually be run on a computer, the human-readable form must be converted to a machine code, (also known as machine-readable code), which means something to the central processing unit (CPU) of the computer.  This process is called either interpretation or compilation depending on the programming language.  

These two highly disparate but deeper levels of categorization we now consider in more detail.  

\subsection*{The basic alphabet of life}
The coding and other functional regions of DNA are \textit{ipso facto} non-random in sequence, but the overall sequence of bases in the genome is also demonstrably non-random, as exemplified by innumerable features, including the following examples. First, there are wide variations in the C+G content of genomes \cite{Mann20107,Li2014}. Second, there is underrepresentation of CG dinucleotide repeats \cite{DeatonBird2011,Yomo1989}. Third, Chargaff's second parity rule notes the occurrence of (approximately) equivalent numbers of nucleotides (G = C; A = T) and of nucleotide motifs with their reverse complement on the same strand \cite{AlbrechtBuehler2006}. Fourth, there are many classes of repeated DNA sequences distributed throughout, and accounting for significant proportions of, the genome \cite{Biscotti2015}. 

Typically these imbalances in DNA sequences are explained by specific mechanisms. For example they have been described as adaptive responses to the environment in prokaryotes  \cite{Mann20107}. In the case of CG dinucleotides, their involvement in controlling gene expression (as CpG islands), their particular secondary structure, and their role as sites for mutation \cite{DeatonBird2011,Kaushik2016} are all considered reasons for their underrepresentation. In the case of repeated DNA, many of these sequences are mobile genetic elements that can be either transposed or copied and transposed in the genome, leading to an accumulation of these elements in the genome. Over 40 percent of the human genome and over 80 percent of the maize genome consist of transposable elements \cite{Tenaillon2011}. 

Thus, the structure of the genome represents the outcome of multiple mechanisms operating over evolutionary timescales - the generation of coding and transcriptional control sequences, the response to diverse biochemical and environmental pressures, and the invasion and spread of transposable elements. However, the predictions of CoHSI transcend all of these essentially local complexities and treat the genome simply as long strings of symbols (to which we attach no meaning or significance). The predictions of CoHSI with respect to the structure of the genome are global - they make no distinction between coding and non-coding, functional or non-functional, and unique or repeated sequences; the whole genome, we predict, is subject to the constraints of CoHSI.  

\subsection*{The basic alphabet of software}
In computers, the most fundamental level of arithmetic has 2 symbols in contrast to the 4 symbols of DNA.  Machine-readable code appears as long strings of 1s and 0s but in practice for display purposes and to compress it, it is shown in hexadecimal format, (i.e. 4 bits at a time), numbered 0-9,a-f, with a-f corresponding to decimal numbers from 10 to 15 respectively.  The hexadecimal character $e$, corresponding to the decimal value 14 is therefore an abbreviated form for a bit pattern of $1110 = 2^{3} + 2^{2} + 2^{1}$, whereas the hexadecimal character $4$ is the same as its decimal form and is an abbreviated form for a bit pattern of $0100 = 2^{2}$.  A typical string of hexadecimal characters from a binary dump on a Linux machine of a computer program looks like

\begin{verbatim}
...
4102301400e08a2625010b32000e0242
...
\end{verbatim}

What these characters mean and where they appear is the function of the systems software of a computer, (c.f. ELF format\footnote{https://en.wikipedia.org/wiki/Executable\_and\_Linkable\_Format, accessed 28-Nov-2018}.)  Some parts contain only data and some parts are control sections which direct the CPU and the computer hardware to do something.  There is a weak analogy with the coding and non-coding regions of DNA, but it would be easy to over-exaggerate this and we will avoid simple but potentially misleading analogies wherever possible.  In spite of their rapidly growing size and ubiquity, computer programs barely even hint at the levels of complexity we observe in the genome  - we understand precisely how coding and data sections in a computer work, but we have an incomplete understanding of how the genome integrates its functions to maintain and propagate life.  We should also note that the mechanisms of life work upwards from the genome to higher-level objects (cells, tissues and organs, organisms) whereas the process of preparing a computer program to run goes in the opposite direction from human cognition to higher-level human readable instructions down to the machine-readable binary format understood by the computer hardware.

We have then two kinds of system of distinct provenance in which we have previously demonstrated CoHSI patterns at the level of categorization that recognizes components \cite{HattonWarr2016,HattonWarr2017,HattonWarr2018a,HattonWarr2018d}; in the case of the genome these components are the encoded proteins (composed of amino acids) and in the case of computer programs the components are computer functions or routines (written in source code). In the study reported here we test the predictions of CoHSI at the lowest level of categorization for these two systems; the four bases of the genome and the hexadecimal format of computer machine-readable code.  We emphasize that this  is an entirely different level of abstraction from the categorization used to study proteins and computer functions. In contrast to the clear distinction made between components in systems of proteins or of computer functions, \textit{here we treat the genome and machine code simply as long uninterrupted strings of symbols, either the bases of DNA or the hexadecimal characters of machine-readable code respectively.}  The only boundaries we will recognize are the start and end of the genome of a species and the start and end of the machine code of a computer program, and even these we will ignore when we examine the patterns predicted in \textit{aggregates} of genomes or of computer programs. Furthermore, we also examine for both proteins and computer software whether the observed structural patterns are independent of reading frame, i.e. the position in the string of symbols at which we start reading, or even how many symbols we include in the reading window, (we consider 1,2,3,4,5,6,7 and 8).  Other authors have considered the tuple occurrence problem.  Specifically, \cite{Gan2011} use a model of preferential attachment as a generator for power-laws and consider very long tuples, whereas in this study we test the predictions of the CoHSI conservation principle.

As we describe in more detail below, CoHSI predicts that in such linear strings of symbols as DNA or machine-readable code, motifs consisting of strings (tuples) of symbols will occur with a frequency, when rank ordered, that corresponds to a power law (i.e. Zipf's Law). 

Although we confine ourselves to the genome and to machine code here, we have previously discussed the application of CoHSI principles to texts \cite{HattonWarr2017} and must therefore point out similar studies. \cite{StephensBialek2010} address the frequency of four-letter words in the collected works of Jane Austen and a mixed set of American-English texts whereas \cite{Piantadosi2014} discusses the semantics of English and the ubiquity of Zipf's law, closing with the statement "To make progress at understanding why language obeys Zipf's law, studies must seek evidence beyond the law itself, testing assumptions and evaluating novel predictions with new, independent data."  This we have done simply by including texts under the umbrella of qualifying discrete systems in which CoHSI operates \cite{HattonWarr2017}.  These papers \cite{StephensBialek2010,HattonWarr2017} add considerable experimental weight to the predictions of the CoHSI models.

\section*{Methods} 
\subsection*{Theoretical background}
Some kind of underpinning theoretical model is essential when approaching datasets of the size of genomes or of computer programs looking for patterns, otherwise it is all too easy to be overwhelmed by their gigantic and growing complexity, not to mention "p-hacking" if the observations are not tested in the context of a clear prediction. The theory leading to CoHSI is described in \cite{HatTSE14,HattonWarr2015} and fully developed in \cite{HattonWarr2017}, but it will be useful to spend a few paragraphs describing exactly how this methodology of CoHSI can be used to approach problems of categorization without any consideration of specific mechanisms.

\subsubsection*{The Classification of Discrete Systems and the Predictions of CoHSI}
In \cite{HattonWarr2017}, we noted that discrete systems  (i.e. systems whose pieces can be counted as integers) can be divided into two distinct types. The indivisible pieces can be considered synonymously as symbols, signs or as tokens, with token being our preferred term in discussing proteins and software. In   \textit{Heterogeneous} systems the indivisible tokens taken from an alphabet of tokens are assembled sequentially, in distinguishable order, into larger units that we term components. The concept of components can be illustrated by considering the proteome (which is the collection of all the proteins expressed by a species) as a discrete system. The components in this case are the proteins, which are themselves assembled sequentially and in distinguishable order from amino acids (the tokens). Another example is provided by the  computer functions that are assembled sequentially and in distinguishable order from computer source code (the tokens) and that constitute the components of a computer program. The second type of discrete system that we distinguish is  \textit{Homogeneous}. Homogeneous discrete systems are simply those in which the indivisible units can be counted without reference to any order of assembly - a classical example of a homogeneous system is the frequency of words (equivalent here to tokens) in a written text. In this example the assignment of words is to bins, where each bin (the equivalent here of a component)  contains only the occurrences of a specific word. When the bins are rank ordered by the frequencies of the words they contain, a power law relationship (the classical Zipf's Law \cite{Zipf35}) emerges.

The above discussion implies that a single discrete system might be categorized as both a heterogeneous and a homogeneous system. CoHSI theory predicts not only different behaviours for heterogeneous and homogeneous systems, but also that both behaviours can co-exist simultaneously in a single system, dependent solely upon the categorization used \cite{HattonWarr2017}. For example in the heterogeneous categorization of a text such as a book the words are components and the letters are tokens and the distribution of word lengths thus follows the predicted canonical CoHSI length distribution \cite{HattonWarr2017} for heterogeneous systems. When the same book is re-categorized as a \textit{homogeneous} system, i.e. when the words themselves are designated as tokens and binned according to their frequency of occurrence, CoHSI predicts that homogeneous systems will be overwhelmingly likely to obey a simple power-law (in fact CoHSI provides a proof of Zipf's law \cite{HattonWarr2017}). The same coincident heterogeneous and homogeneous behaviour is not unique to texts - it was also observed in the currently known set of proteins \cite{HattonWarr2018d}, whereby on one hand the length distribution of proteins measured in amino acids follows the canonical heterogeneous form while on the other hand, the distribution of frequency of protein re-use (seen when gene transfer has occurred between species) is power-law by rank as predicted for the homogeneous categorization.  

%
%

The full development of CoHSI theory is in \cite{HattonWarr2017}, and for the homogeneous case for reasonably large values of the contents of the $i^{th}$ component $t_{i}$, (where Stirling's approximation is valid), the distribution is the solution of

\begin{equation}
\log t_{i} = -\alpha -\beta ( \frac{d}{dt_{i}} \log N(t_{i}, a_{i}; a_{i} ) )    \label{eq:minifst}
\end{equation}

The derivation of this uses the simplest form of Stirling's approximation $\log (t_{i} !) \approx t_{i} \log t_{i} - t_{i}$ and in the homogeneous case, the information function $N(t_{i}, a_{i}; a_{i} )$ of \cite{HattonWarr2017}, is simply $r^{t_{r}}$, where $r$ is the rank order and so the frequency distribution is given by

\begin{equation}
\log t_{r} = -\alpha -\beta ( \frac{d}{dt_{r}} \log r^{t_{r}} ) =  -\alpha -\beta \log r   \label{eq:minifstr}
\end{equation}

The solution of this is simply Zipf's power-law for all ranks \cite{HattonWarr2017}. 

\subsection*{Slicing and Dicing the Genome}
We analyzed the genomes of the ten species shown in Table \ref{tab:genomes}, downloaded from various public sites (the Sanger Institute and the NCBI). These genomes varied in size from 0.8 MB (Megabyte) to 881.2 MB and from approximately 10\% to 98\% non-protein coding sequence.  We have preserved the file names to identify uniquely the particular assemblies. The sequences were analyzed for non-overlapping n-tuples (n=1-8) read sequentially from the beginning of the sequence to the end, except where offsets (changes in reading frame) are specifically noted. We did not use pre-existing databases of tuple sequences - the bins were defined by the tuple sequences that emerged from the analysis.

\begin{table}[!ht]
\centering
\begin{tabular}{p{4cm}rc}
\hline
Species & Genome file & size (MB.)\\
\hline
Ciona & Ciona\_intestinalis.KH.dna.toplevel.fa.gz & 34.4 \\
Fruitfly & Drosophila\_melanogaster.BDGP6.dna.toplevel.fa.gz & 42.3 \\
Rice & GCA\_001889745.1\_Rice\_IR8\_v1.7\_genomic.fna.gz & 122.6 \\
Cyanobacteria & GCA\_003555505.1\_ASM355550v1\_genomic.fna.gz & 0.8 \\
Mushroom & GCA\_000300555.1\_Agabi\_varbur\_1\_genomic.fna.gz & 9.7 \\
Yeast & Saccharomyces\_cerevisiae.R64-1-1.dna.toplevel.fa.gz & 3.8 \\
Nematode & Caenorhabditis\_elegans.WBcel235.dna.toplevel.fa.gz & 30.3 \\
Human & Homo\_sapiens.GRCh38.dna.primary\_assembly.fa.gz & 881.2 \\
Gorilla & Gorilla\_gorilla.gorGor4.dna.toplevel.fa.gz & 873.4 \\
Thale cress & GCA\_000001735.2\_TAIR10.1\_genomic.fna.gz & 37.5 \\
\hline
\end{tabular}
\caption{The ten genomes used in our study and their sources. The mushroom sequence is of \textit{Agaricus bisporus} and the cyanobacterial sequence is of \textit{Thermosynechococcus elongatus} PKUAC-SCTE542.}
\label{tab:genomes}
\end{table}

\section*{Results and Discussion}

\subsection*{Testing Predictions Using Ten Genome Sequences}

Our straightforward prediction using homogeneous CoHSI theory is that the occurrence rate of n-tuples will follow a power-law in rank order of occurrence, i.e. Zipf's law, for any n.

\begin{figure}[ht!]
\centering
\includegraphics[width=12cm]{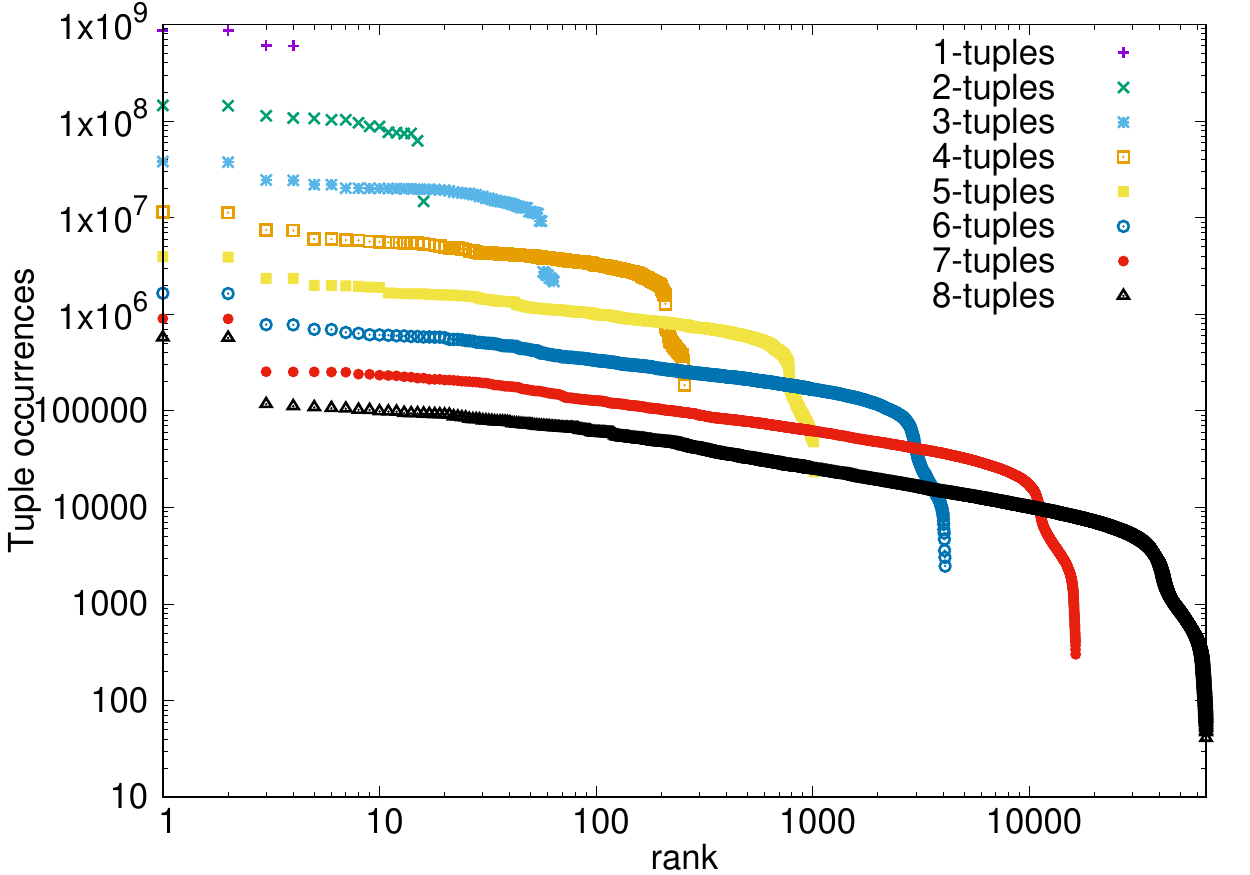}
\caption{The occurrence rates of n-tuples in  the human genome for $n$ = 1, 2, 3, 4, 5, 6, 7 and 8. The plots are of the complementary cumulative distribution function (ccdf)}
\label{fig:human}
\end{figure}

First we attempted to falsify this prediction by analyzing the haploid human genome as cited in Table \ref{tab:genomes} for $n$ = 1, 2, 3, 4, 5, 6, 7 and 8.  The results are shown in Fig. \ref{fig:human}; the presence of the predicted power-law (linearity on a log-log plot) is clear and seen emphatically with the higher tuples, where the number of bins has a potential maximum of 65,536 in the case of the 8-tuples. 

\textit{R lm() reports that the associated p-value matching the power-law linearity in the 8-tuple  ccdf of Fig. \ref{fig:human} is $< (2.2) \times 10^{-16}$ over the 3-decade range $20-20000.0$, with an adjusted R-squared value of $0.997$.  The slope is $-0.41 \pm 0.02$.}

The same analysis of 1-8 tuples was carried out with the other 9 genomes (Appendix Figs 12 and 13), and in every case power law distributions of tuple frequency (as observed with the human genome) were seen regardless of species.

The following points can be noted

\begin{figure}[ht!]
\centering
\includegraphics[width=14cm]{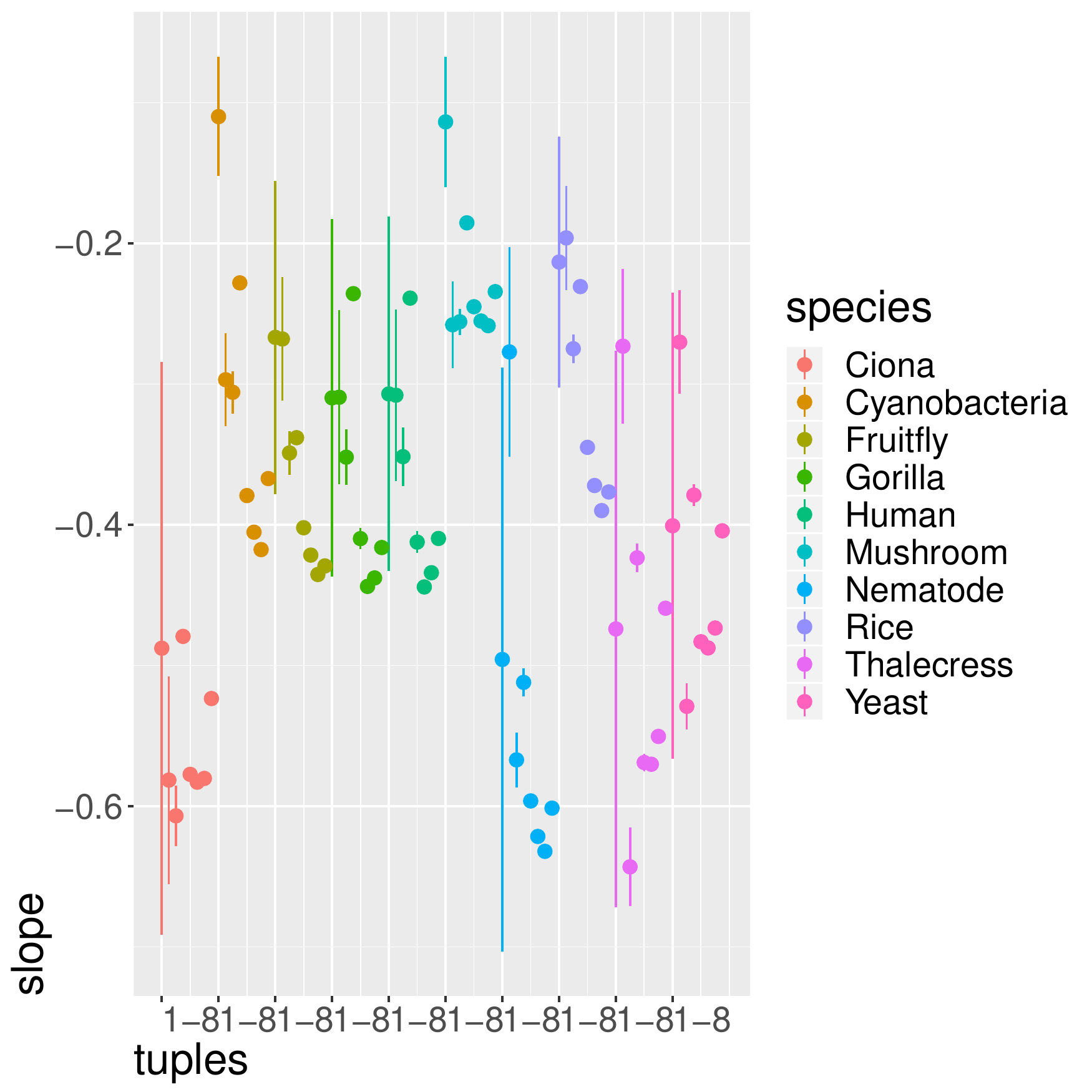}
\caption{Each slope with error bounds extracted from R analysis of each tuple for each species.  The data are shown in tuple order (1-8) for each species.}
\label{fig:allslopes}
\end{figure}

\begin{figure}[ht!]
\centering
\includegraphics[width=12cm]{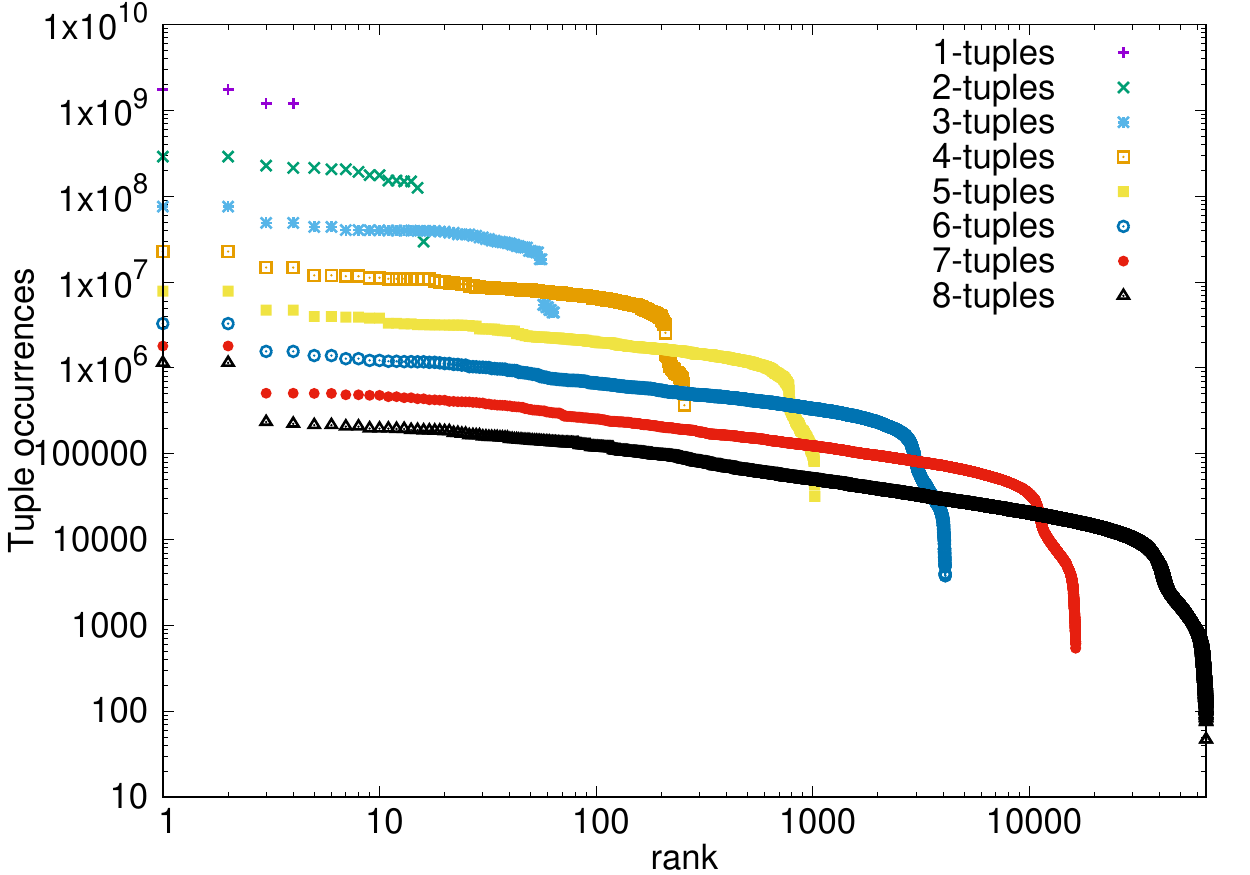}
\caption{The occurrence rates of n-tuples in the human genome, with the forward and reverse sequences of the haploid genome concatenated, for $n$ = 1, 2, 3, 4, 5, 6, 7 and 8. Data are plotted as the complementary cumulative distribution function.}
\label{fig:human2}
\end{figure}

\begin{itemize}
\item All 10 genomes showed identical power law (Zipfian) behaviour for all n-tuples regardless of  variation in the size of the genomes (which spans more than 3 decades between the cyanobacterium and human)  or the proportion of coding sequence they contain, which ranges from 1-2\% in human and gorilla to 20-30\% for the fruitfly, rice, thale cress \textit{Ciona} and \textit{C. elegans}, to approximately 65 - 85\% in the two fungal species (yeast and mushroom) and to 90\% in the cyanobacterium. 
\item The rapid fall-off (droop) at the extreme tail was observed in all genomes and all tuples from 2-8 and occurs in the region where the rarer tuples are not statistically well-represented. Although the drooping tail represents only a small fraction of the genome (<1\%),  nevertheless it does not appear random, and we will discuss the potential origins of this feature of the tail in more detail below.
\item In all species there is a significant departure from CoHSI in the highest ranked tuples (typically the first and second but sometimes extending to additional ranks) where they occur much more frequently than predicted. In the cyanobacterium this deviation is apparent only for the 6-, 7- and 8-tuple analyses. This observation will also be discussed below.
\item The scale independence.  The power-law slopes are almost parallel for all n-tuples in the same genome as shown in Fig. \ref{fig:allslopes}.  To see the similarity from a different viewpoint, the 1-8 tuples for each species we studied are shown individually in the Appendix Figs 12 and 13.
\item Since CoHSI is token-agnostic, we predicted that the result would be unaffected, (apart from a two-fold scale change), when we generated the reverse complementary strand (using $A \rightleftharpoons T$ and $C \rightleftharpoons G$, and reading 5' to 3') and concatenated it with the first strand. We show that this is indeed the case for the human genome in Fig. \ref{fig:human2}.
\end{itemize}

\subsubsection*{The droopy tail}
\begin{figure}[ht!]
\centering
\includegraphics[width=12cm]{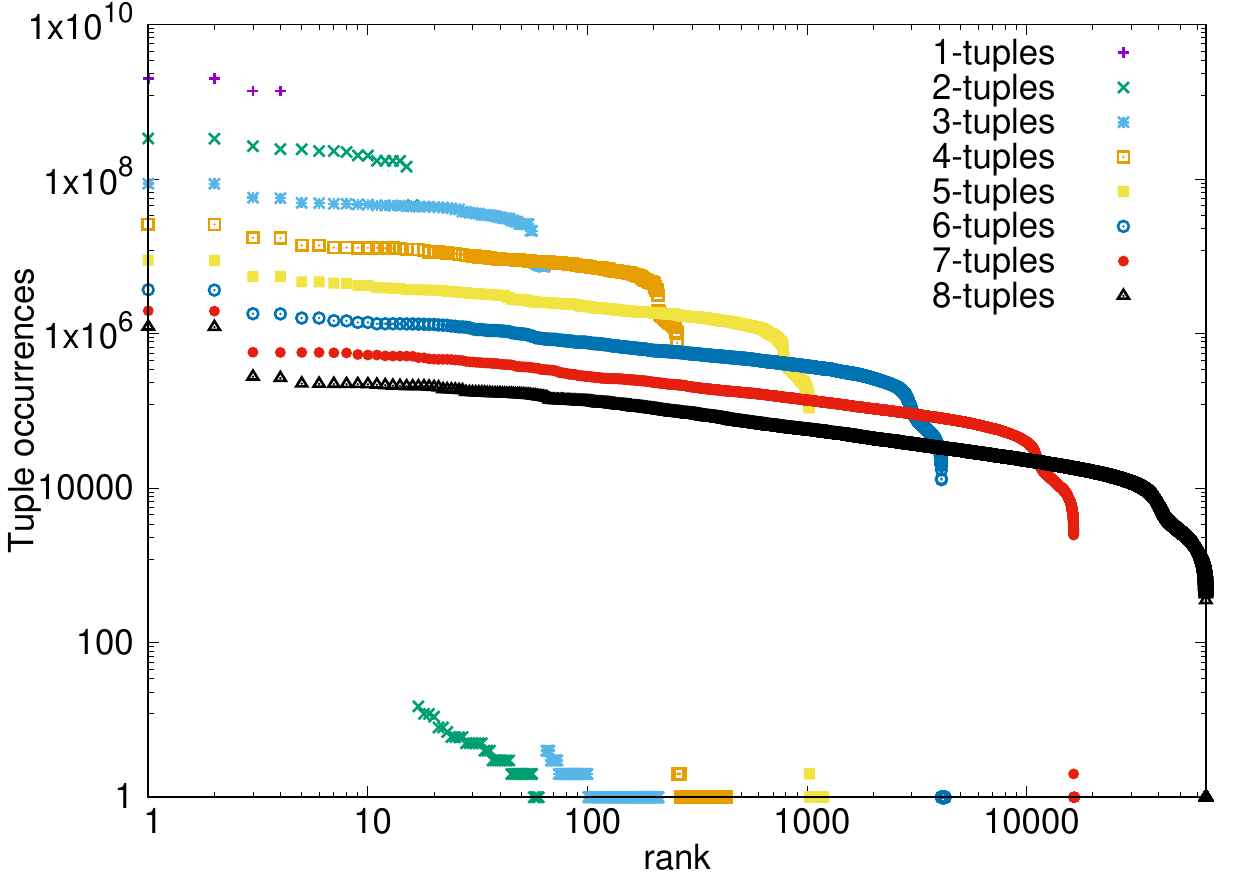}
\caption{The occurrence rates of n-tuples in the combined genomes of all ten species for n = 1, 2, 3, 4, 5, 6, 7 and 8. The anomalous very low frequency bins that are evident close to the x-axis after a break in the drooping tail arise from ambiguous sequence reads in the Thale cress (see Table \ref{tab:bigsmall} and Appendix Fig 13). Data are plotted as the complementary cumulative distribution function. }
\label{fig:combined}
\end{figure}

Initially, we thought the droopy tail was simply a statistical phenomenon caused by the lowest ranked bins having insufficient occupancy, something also noted by \cite{Gan2011}.  It is however not random.  This can be seen by combining the data for the genomes of all ten species we considered into a single large file that was then scanned in one pass from beginning to end as shown in Figure \ref{fig:combined}. The droopy tail is plainly evident and with a similar qualitative nature to that seen in each of the individual species.  This systematic nature of the droopy tail of the homogeneous distribution initially concerned us as CoHSI predicts a pure Zipf's law as we have seen above.  However, we then realized that the pure Zipf's law predicted by CoHSI arises directly from the use of Stirling's approximation.  Wherever Stirling's approximation is good, which is for most of the distribution, a near linear power-law in rank is found as expected.  However for the lowest ranked (i.e. least frequently occurring) items, the basic assumption underlying Stirling's approximation of well-occupied categories no longer holds.

What then happens to the predicted Zipf's law for the least frequently occurring items if Stirling's approximation is not used?  We explored this first using Ramanujan's form \cite{Ramanujan1988} for a better approximation than Stirling to $\log (t_{r} !)$, and this is given by

\begin{equation}
\log t_{r} + \frac{1 + 8 t_{r} + 24t_{r}^{2}}{6(t_{r} + 4 t_{r}^{2} + 8 t_{r}^{3})} = -\alpha -\beta \log r    \label{eq:minif}
\end{equation}

\begin{figure}[ht!]
\centering
\includegraphics[width=12cm]{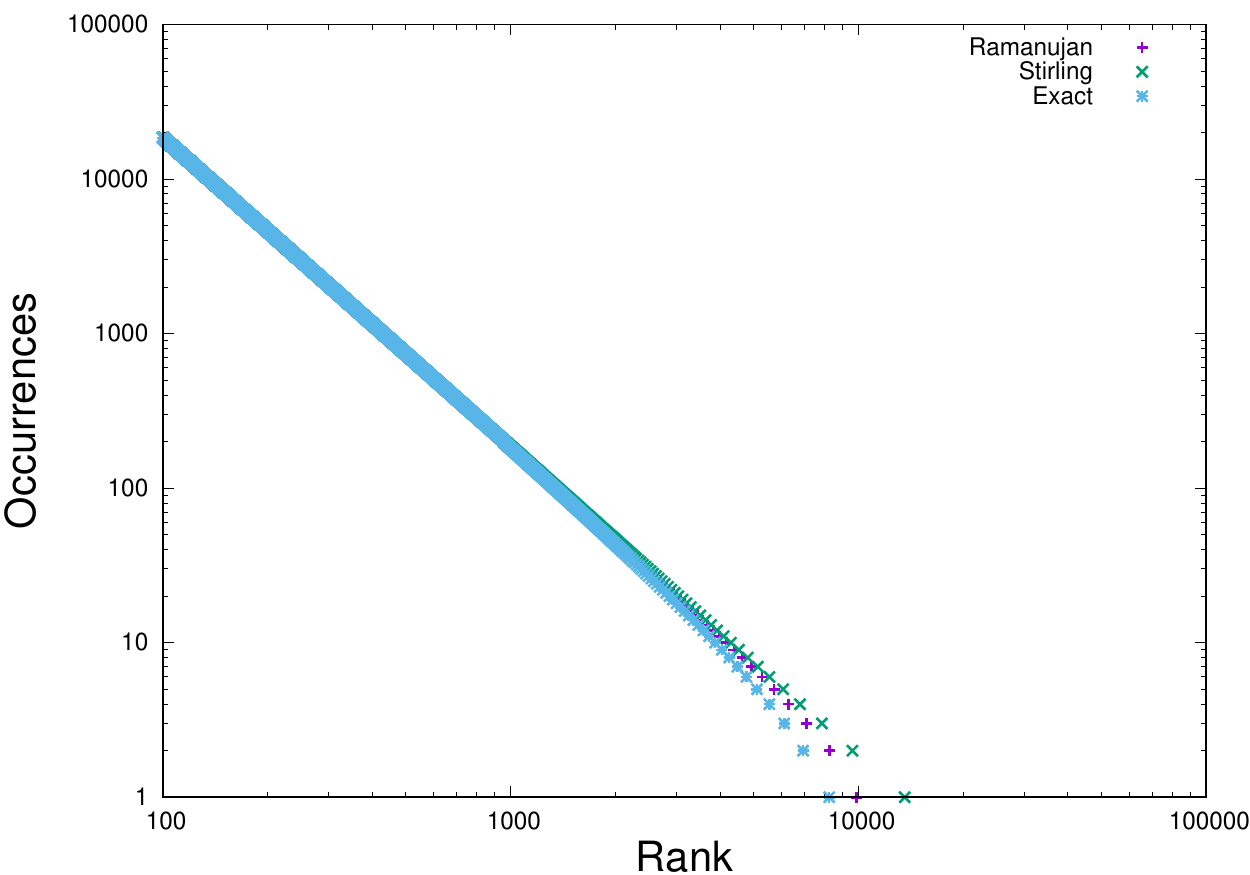}
\caption{Illustrating the droop: the departures from the exact Zipf's law prediction of CoHSI as the $t_{r}$ (occupancy rates of the bins) become small (in essence, this is the y-axis) using Stirling's approximation (which gives exactly Zipf's law), Ramanujan's approximation which begins to depart from Zipf's law, and an exact computation which shows a greater departure when the $t_{r}$ are smallest. The graph illustrates a homogeneous distribution with alpha = -0.001, beta = 2.0}
\label{fig:exact}
\end{figure}

This becomes identical to Stirling's approximation for the highest occupancy ranks as can be seen by letting $t_{r} \rightarrow \infty$ in (\ref{eq:minif}) and comparing with (\ref{eq:minifstr}), but how does it depart from Zipf's law as the occupancy of ranks falls and they appear later in the rank-ordering?  The result can be seen in Fig. \ref{fig:exact}, where a droop begins to appear.  Encouraged by this, we therefore supplemented the Stirling and Ramanujan approximations by computing $\log (t_{r} !)$ \textit{exactly} as $\sum_{k=1}^{k=t_{r}} \log k$ when $t_{r}$ is below some suitable value, (100 in our studies).  The droop increases and is not therefore an artifact of the approximation we used.  This process reveals that the correct CoHSI prediction is that the Zipf's law relation for rank ordering will naturally break down (droop) for the highest values of the rank (i.e. the least occupied bins) as shown in Fig. \ref{fig:exact}.

However, this is clearly not the whole story as the droop is systematic in the tail even when bins in this region are relatively well-occupied and it is observable in all genomes regardless of scale in a qualitatively identical manner.  It seems likely that some systematic departure is present which causes a departure from the equilibrium power-law and the modest droop predicted by CoHSI; further investigation is necessary.

\subsubsection*{Over-Representation Of The Most Frequent Tuples}
 Table \ref{tab:bigsmall} gives the four most common and four least common 8-tuples for each of the 10 genomes analyzed, and in all species except \textit{Ciona}, the two most highly represented 8-tuples occur more frequently than predicted by CoHSI (see Figures 1 and 2, the analyses of n-tuple frequencies for all 10 genomes).  Several points are notable. First, in all 9 of the eukaryotic species the motifs (A)${_8}$ and (T)${_8}$ are always amongst the top 4. Second, the results support Chargaff's Second Parity Rule, even for octanucleotide motifs, in all 10 of the species examined; the gorilla is the only eukaryotic species in which the top four 8-tuples do not constitute two pairs of reverse complementary sequences.  We will address below the manner in which CoHSI provides proof of Chargaff's Second Parity Rule. Third, the cyanobacterial sequence is interesting in that the most frequent 8-tuple is its own reverse complement, with the next two most frequent 8-tuples forming a reverse complementary pair. Fourth, in the nine eukaryotic species the most common 8-tuple motifs do not occur at random, being shared between genomes. Of the 20 most frequent motifs observed when all genomes combined are analysed (Table \ref{tab:mostcommon}) all of them occur in more than one species, being represented in a number of eukaryotic genomes ranging from 2-9. This sharing of common motifs did not extend to the prokaryote - none of these motifs (Table \ref{tab:mostcommon}) was shared with the cyanobacterial genome. 

\begin{table}[!ht]
\centering
\begin{tabular}{p{4cm}ll}
\hline
Species & Most common & Least common \\
\hline
\texttt{Ciona} & \texttt{ATATATAT (7191)} & \texttt{GACGGGCT (7)} \\
     & \texttt{TTTTTTTT (7156)} & \texttt{CCCCCGGA (8)} \\
     & \texttt{AAAAAAAA (7062)} & \texttt{CCTAGCCG (8)} \\
     & \texttt{TATATATA (7014)} & \texttt{CGGGGCGC (8)} \\
\texttt{Cyanobacteria} & \texttt{GCGATCGC (509)} & \texttt{AAAACGAC (1)} \\
     & \texttt{CGATCGCC (321)} & \texttt{AAAACGCA (1)} \\
     & \texttt{GGCGATCG (284)} & \texttt{AAAAGAAC (1)} \\
     & \texttt{GATCGCCC (147)} & \texttt{AAACAAGG (1)} \\
\texttt{Fruit Fly} & \texttt{TTTTTTTT (16876)} & \texttt{CCTAGGGG (24)} \\
     & \texttt{AAAAAAAA (16803)} & \texttt{CGCTAGGG (30)} \\
     & \texttt{ATATATAT (6853)} & \texttt{GGTACCCG (30)} \\
     & \texttt{TATATATA (6137)} & \texttt{CCCCTACG (32)} \\
\texttt{Gorilla} & \texttt{TTTTTTTT (559235)} & \texttt{CGCGTACG (29)} \\
     & \texttt{AAAAAAAA (550006)} & \texttt{CGTCGACG (30)} \\
     & \texttt{ATTCCATT (146972)} & \texttt{GTCGATCG (31)} \\
     & \texttt{TTCCATTC (143536)} & \texttt{CGTACGCG (34)} \\
\texttt{Human} & \texttt{TTTTTTTT (573882)} & \texttt{CGTACGCG (22)} \\
     & \texttt{AAAAAAAA (570162)} & \texttt{CGCGTACG (25)} \\
     & \texttt{ATATATAT (117335)} & \texttt{TCGCGTCG (31)} \\
     & \texttt{TATATATA (111608)} & \texttt{CGCGTCGA (34)} \\
\texttt{Mushroom} & \texttt{TTTTTTTT (1115)} & \texttt{GGGCCCCT (4)} \\
     & \texttt{AAAAAAAA (936)} & \texttt{CCCCTAAG (7)} \\
     & \texttt{GAAGAAGA (444)} & \texttt{GCCCCTAG (7)} \\
     & \texttt{TCTTCTTC (419)} & \texttt{GGGGCCCG (7)} \\
\texttt{Nematode} & \texttt{TTTTTTTT (18378)} & \texttt{CGCCCGGG (8)} \\
     & \texttt{AAAAAAAA (18372)} & \texttt{GGGCCGCT (8)} \\
     & \texttt{ATTTTTTT (9642)} & \texttt{GGGGCCCT (8)} \\
     & \texttt{AAAAAAAT (9621)} & \texttt{CTAAGGGG (9)} \\
\texttt{Rice} & \texttt{TATATATA (38933)} & \texttt{CGCGCTTA (63)} \\
     & \texttt{ATATATAT (38420)} & \texttt{GCGTTACG (65)} \\
     & \texttt{AAAAAAA (30448)} & \texttt{CGTAACGC (67)} \\
     & \texttt{TTTTTTT (30367)} & \texttt{CGCGTTAC (79)} \\
\texttt{ThaleCress} & \texttt{AAAAAAAA (16120)} & \texttt{AAACKAGA (1)} \\
     & \texttt{TTTTTTTT (15921)} & \texttt{AAAGMMWC (1)} \\
     & \texttt{ATATATAT (7491)} & \texttt{AAAMCCTA (1)} \\
     & \texttt{TATATATA (7199)} & \texttt{AACAWWWT (1)} \\
\texttt{Yeast} & \texttt{AAAAAAAA (1172)} & \texttt{AACACGCG (1)} \\
     & \texttt{TTTTTTTT (1099)} & \texttt{AACCGCGC (1)} \\
     & \texttt{ATATATAT (536)} & \texttt{ACACGCGT (1)} \\
     & \texttt{TATATATA (506)} & \texttt{ACACGGTC (1)} \\
\hline
\end{tabular}
\caption{The four most and least common 8-tuples in each species, with their frequencies of occurrence in parentheses. We note that the 4 least frequent motifs in Thale Cress have all arisen artefactually from ambiguous sequence reads}
\label{tab:bigsmall}
\end{table}

\begin{table}[!ht]
\centering
\begin{tabular}{cccc}
\hline
Rank & Motif & Frequency & \# genomes \\
\hline
1 & TTTTTTTT & 1224036 & 9 \\
2 & AAAAAAAA & 1211089 & 9 \\
3 & ATATATAT & 277004 & 7 \\
4 & TATATATA & 264803 & 6 \\
5 & ATTTTTTT & 225262 & 7 \\
6 & AAAAAAAT & 223365 & 7 \\
7 & TTTTAAAA & 222933 & 3 \\
8 & TTTATTTT & 221810 & 5 \\
9 & AAAATAAA & 221143 & 6 \\
10 & TGTGTGTG & 220136 & 3 \\
11 & CACACACA & 217608 & 3 \\
12 & TTTCTTTT & 214220 & 3 \\
13 & GTGTGTGT & 211654 & 3 \\
14 & AAATAAAA & 209876 & 4 \\
15 & TTTTATTT & 208341 & 4 \\
16 & TTTTTAAA & 206897 & 3 \\
17 & ACACACAC & 206833 & 3 \\
18 & TTTAAAAA & 206671 & 2 \\
19 & TTTTCTTT & 204638 & 4 \\
20 & AAAGAAAA & 202723 & 4 \\
\hline
\end{tabular}
\caption{The most frequent 8-tuples in all 10 genomes combined, their frequency, and the number of genomes in which they occur in Table \ref{tab:bigsmall}.}
\label{tab:mostcommon}
\end{table}

\subsubsection*{Sensitivity analysis}
CoHSI is a token-agnostic and mechanism-agnostic theory.  We would therefore expect that the emphatic power-law behaviour for the n-tuples reported above would be independent of the start of the reading frame.  This was tested on all genomes for the largest tuple considered, the 8-tuple.  The results (data not shown) were identical with no sensitivity noted for start offsets of 1, 2, 3 and 4 bases.  A typical example is shown for the human genome in Figure \ref{fig:8-4}.  The curves for the 4 analyses are so similar that they are overlain perfectly except for a small number of very minor differences at the extreme ends of the distribution.

It is certainly correct that starting the tuple analysis with non-zero offsets of the reading frame could identify different orderings of tuples, but we must recall that the homogeneous model is in \textit{rank} order.  CoHSI simply leads in each case to a re-ordered list which, when arranged by rank leads to precisely the same predicted power-law.  It is worth noting that this result is apparently independent of any distinction between protein-coding and non-protein coding regions of the genome. The start codon for any protein could randomly be at an offset of 0, 1 or 2 from the tuple reading frame (or even in 3'-5' orientation because it is read from the second strand), the effects of which would tend to obscure any contribution to the results of protein reading frame. However, it is perhaps more important to recall that estimates for the non-protein coding proportion of the genome range from around 17\% for yeast to as high as 98\% for primates, and the results that we observe are qualitatively identical for all 10 genomes considered individually or in aggregate - i.e. as CoHSI predicts, the proportion of non-protein coding sequence in a genome has no effect on the distribution of tuples.  

\begin{figure}[ht!]
\centering
\includegraphics[width=12cm]{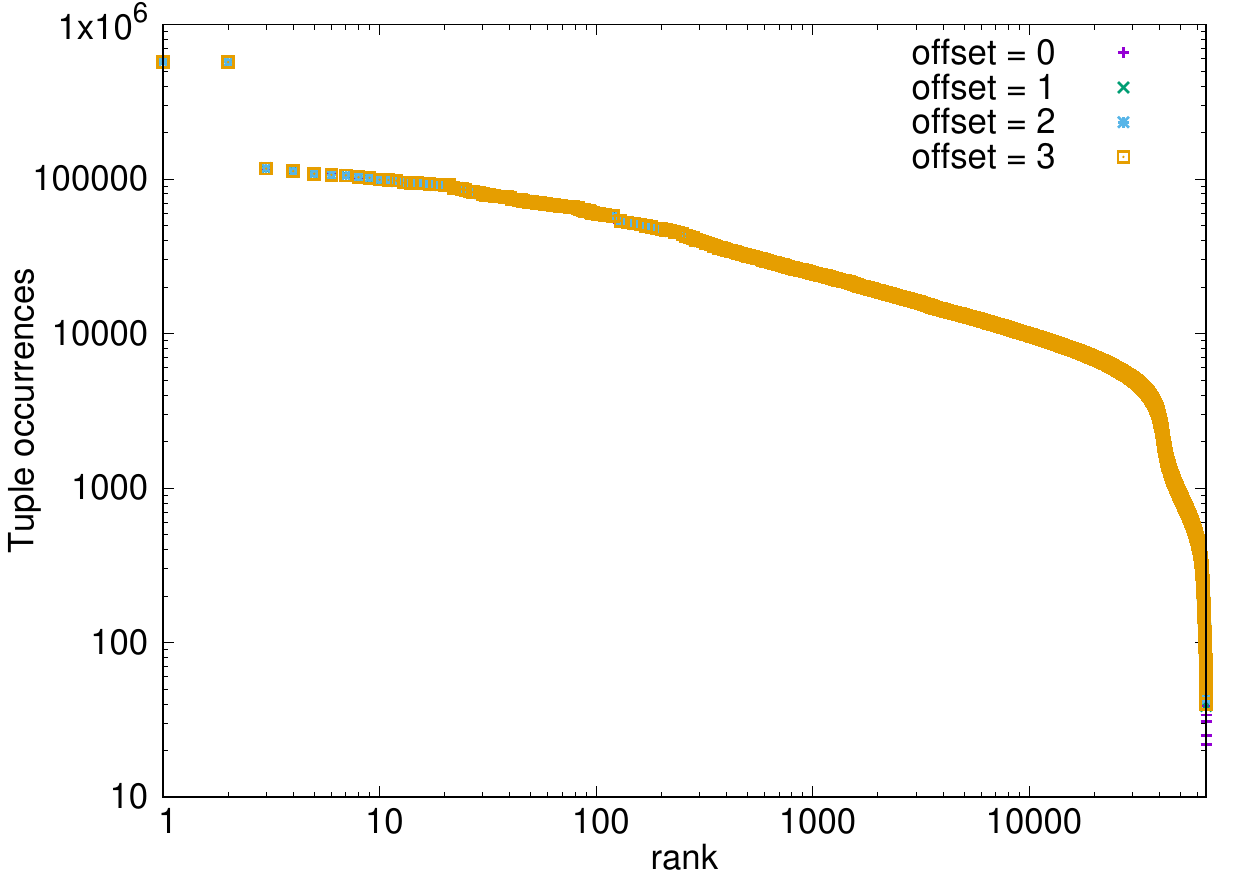}
\caption{The occurrence rates of 8-tuples in the Human genome for offsets of 0, 1, 2 and 3. The 4 curves overlie one another almost exactly.  Note that this graph has been decimated to reduce its size.  Only every 10th point is shown for offsets = 1, 2 and 3 after the first 100 points. Data are plotted as the complementary cumulative distribution function.}
\label{fig:8-4}
\end{figure}

\subsubsection*{Proof of Chargaff's Second Parity Rule}

It can be noted that the analysis of the 20 most frequent motifs that occur in all 10 genomes combined (Table \ref{tab:mostcommon}) includes 8 pairs of motifs that are reverse complements of one another, and 3 motifs that are reverse complements of themselves (dyadic repeats). We extended this analysis to the top 50 most common motifs in all 10 genomes combined (data not shown) and observed that 44 of these motifs form 22 reverse complementary pairs, in addition to the 3  above-noted motifs that are dyadic repeats (ATATATAT, TATATATA and TTTTAAAA). This result was unexpected - it suggests that Chargaff's Second Parity Rule might extend beyond a single genome to larger aggregates of genomes. We calculated the frequency ratios for each of these top 22 pairs of motifs in the 10 genomes combined with the following result. Seven of these pairs showed a ratio of 1.00; ten pairs showed a ratio of 1.01; four pairs showed a ratio of 1.02 and one pair had a ratio of 1.07. Thus these 8-tuple motifs show the clear hallmarks of Chargaff's Second Parity Law; reverse complementary sequences and almost identical frequency of occurrence on a single strand of the DNA. Although our analysis was less than fully comprehensive of all the 8-tuple motifs (a daunting task), the fact that the examined motifs are distributed across multiple genomes (Table \ref{tab:mostcommon}) already suggests that Chargaff's Second Parity Rule may embrace aspects of the global properties of genomes. This prompted us to examine if one of the predictions of CoHSI (which as we have shown above constrains the global properties of genomes) might include Chargaff's Second Parity Rule. The results of this investigation are outlined below.   

Chargaff's rules are empirical and refer to the relative frequencies of A, T, G, and C bases in DNA, which we denote as \%A, \%T, \%C and \%G.  The first rule holds that \%A = \%T and \%G = \%C \textit{globally} in the double-stranded DNA.  This observation, which predated Watson and Crick's ground-breaking discovery, helped guide them in uncovering the structure of the double helix.  Once the structure of the DNA double helix is known, the first of Chargaff's rules is self-evident.

However, the second empirical rule holds that \%A = \%T and \%G = \%C \textit{individually} on each strand of a double-stranded DNA molecule.  It is observed\footnote{https://en.wikipedia.org/wiki/Chargaff\%27s\_rules} that it does not apply to single-stranded DNA or any RNA, or relatively short DNA sequences.  Our work with CoHSI allows us to present a proof of Chargaff's Second Parity Rule.

We suppose that there are four bases W, X, Y and Z with complementarity rules $W \rightleftharpoons X$ and $Y \rightleftharpoons Z$. 

In any assembly, the CoHSI Homogeneous equation states that their frequencies will obey a power-law.  Considering strand 1 of a double-helix pair, the frequency of occurrence of 4 bases in 2 complementary pairs can be written down in $(4!/2!2!) = 6$ unique ways.

\begin{equation}
    \%W \geq \%X \geq \%Y \geq \%Z
    \label{eq:I}
\end{equation}

\begin{equation}
    \%W \geq \%Y \geq \%X \geq \%Z
    \label{eq:II}
\end{equation}

\begin{equation}
    \%W \geq \%Y \geq \%Z \geq \%X
    \label{eq:III}
\end{equation}

\begin{equation}
    \%X \geq \%W \geq \%Y \geq \%Z
    \label{eq:IV}
\end{equation}

\begin{equation}
    \%X \geq \%Y \geq \%W \geq \%Z
    \label{eq:V}
\end{equation}

\begin{equation}
    \%X \geq \%Y \geq \%Z \geq \%W
    \label{eq:VI}
\end{equation}

Consider (\ref{eq:I}).  If we apply complementarity to this, we get for strand 2

\begin{equation}
    \%X \geq \%W \geq \%Z \geq \%Y
    \label{eq:cI}
\end{equation}

However, the CoHSI homogeneous law applies to any significantly sized assembly, so strand 2 must have the same distribution as strand 1.

\begin{equation}
    \%W \geq \%X \geq \%Y \geq \%Z
    \label{eq:I2}
\end{equation}

Combining (\ref{eq:cI}) and (\ref{eq:I2}), the only solution for strand 2 (and therefore strand 1 by symmetry)

\begin{equation}
    \%W = \%X \geq \%Y = \%Z
    \label{eq:sI}
\end{equation}

So for (\ref{eq:I}), CoHSI and complementarity jointly imply that four bases will appear in one strand in two pairs with each pair separated in frequency by a power-law (\ref{eq:sI}).

Consider now (\ref{eq:II}).  If we apply complementarity to this, we get for strand 2

\begin{equation}
    \%W \geq \%Y \geq \%X \geq \%Z
    \label{eq:cII}
\end{equation}

However, applying the CoHSI homogeneous law, again strand 2 must have the same distribution as strand 1.

\begin{equation}
    \%X \geq \%Z \geq \%W \geq \%Y
    \label{eq:II2}
\end{equation}

Combining (\ref{eq:cII}) and (\ref{eq:II2}), the only solution for strand 2 (and therefore strand 1 by symmetry) is then

\begin{equation}
    \%W = \%X = \%Y = \%Z
    \label{eq:sII}
\end{equation}

Continuing like this, we find that (\ref{eq:I}), (\ref{eq:IV}) allow a two complementary pair solution, whereas (\ref{eq:II}), (\ref{eq:III}), (\ref{eq:V}) and (\ref{eq:VI}) only allow a solution where all four bases occur with the same frequency, but this breaks CoHSI since they must distribute as a power-law, implying that the only solution is two complementary pairs on each strand as represented by (\ref{eq:sI}).

We note

\begin{enumerate}
    \item In the Genome, these pairs are W = T, X = A and Y = C and Z = G.
    \item We note that for haploid strands for which no complementarity applies, (\ref{eq:I}) must hold, but not (\ref{eq:cI}) as paired bases need not occur.  The CoHSI power-law will hold however explaining why single-stranded RNA amongst others for example does not obey Chargaff's Second Parity rule.
    \item For smaller assemblies, CoHSI is an asymptotic approximation, consequently we can expect this proof to break down for haploid strands and also for smaller diploid strands.
    \item Finally, CoHSI knows nothing of tokens so it cannot say that any particular complementary pair will be jointly dominant over the other complementary pair.  All it can say is that if there are four bases in a diploid strand with complementarity in force, they will manifest themselves as two pairs of complementary bases with a pair ratio obeying a power-law (there are only two points but the slope will be essentially the same as for for the higher tuples, see Fig 2), or as four identically-occurring bases.  The latter does not conform to a power law relationship and is thus ruled out by CoHSI.
\end{enumerate}

We close this section by showing in Fig. \ref{fig:8-4}, the clear two-pairs structure for the 1-tuples in all species tested.

\begin{figure}[ht!]
\centering
\includegraphics[width=12cm]{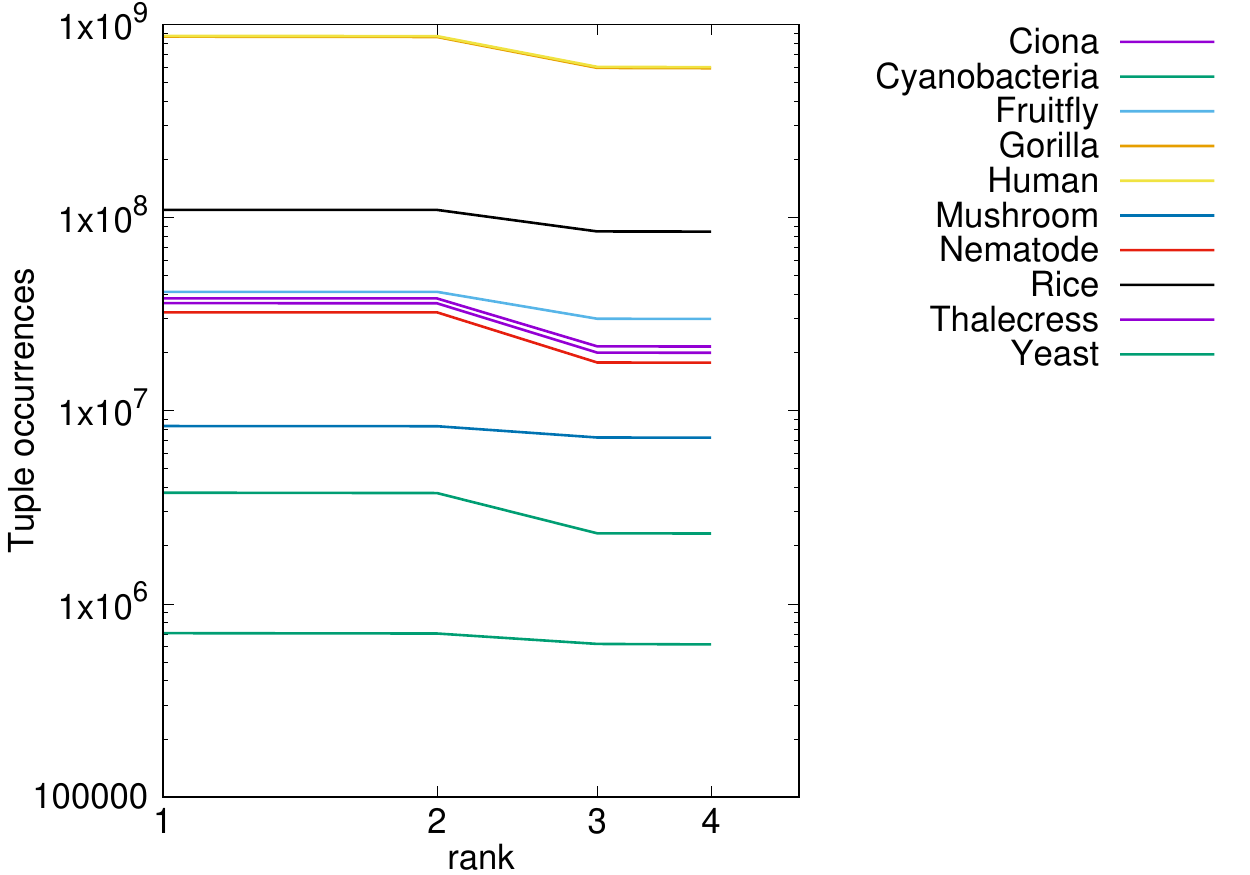}
\caption{The occurrence rates of 1-tuples in all species.  The clear two-pair structure whereby $\%A=\%T$ and $\%G=\%C$ is visible in each species although the differentiation differs.}
\label{fig:8-4}
\end{figure}

\subsubsection*{Extended Chargaff-like symmetries}
Chargaff's Second Parity Rule was originally framed in terms of the frequencies of the individual four bases, and the proof presented above can be extended to the symmetries of higher-tuple motifs in diploid genomes subject to complementarity.

The nature of the above proof shows that \textit{if an n-tuple and its complement are the most commonly occurring in a haploid strand, then CoHSI implies that their frequency will be identical within statistical uncertainty} as they will isolate themselves as a pairing as was the case with 1-tuples in (\ref{eq:sI}).  This is abundantly obvious in Table \ref{tab:bigsmall} with poly(A) and its complement poly(T).  It can also be seen in all tuples from 1 to 8 in Fig \ref{fig:combined} and in Table \ref{tab:mostcommon}.

More intriguing is that \textit{if two n-tuples and their complements happen to appear as the four highest frequencies of occurrence, then CoHSI implies that they can isolate themselves as a 4-group which have the same frequency within statistical uncertainty}.  This is actually visible in the 8-tuple data of Table \ref{tab:bigsmall} for \textit{Ciona}.

Following the proof above for 1-tuples, this line of reasoning with respect to Chargaff's Second Parity Rule and the global properties of motifs within and across genomes could clearly be taken much further. However, when the sample size is just 10 genomes there is sufficient statistical noise to make it challenging to undertake such a comprehensive study. Regardless of this, we consider that the theory and examples presented here clearly point to both an explanation of Chargaff's Second Parity Rule and to its wider application in understanding the structure of genomes.

\subsection*{Testing Predictions using Computer Software}
\begin{figure}[ht!]
\centering
\includegraphics[width=12cm]{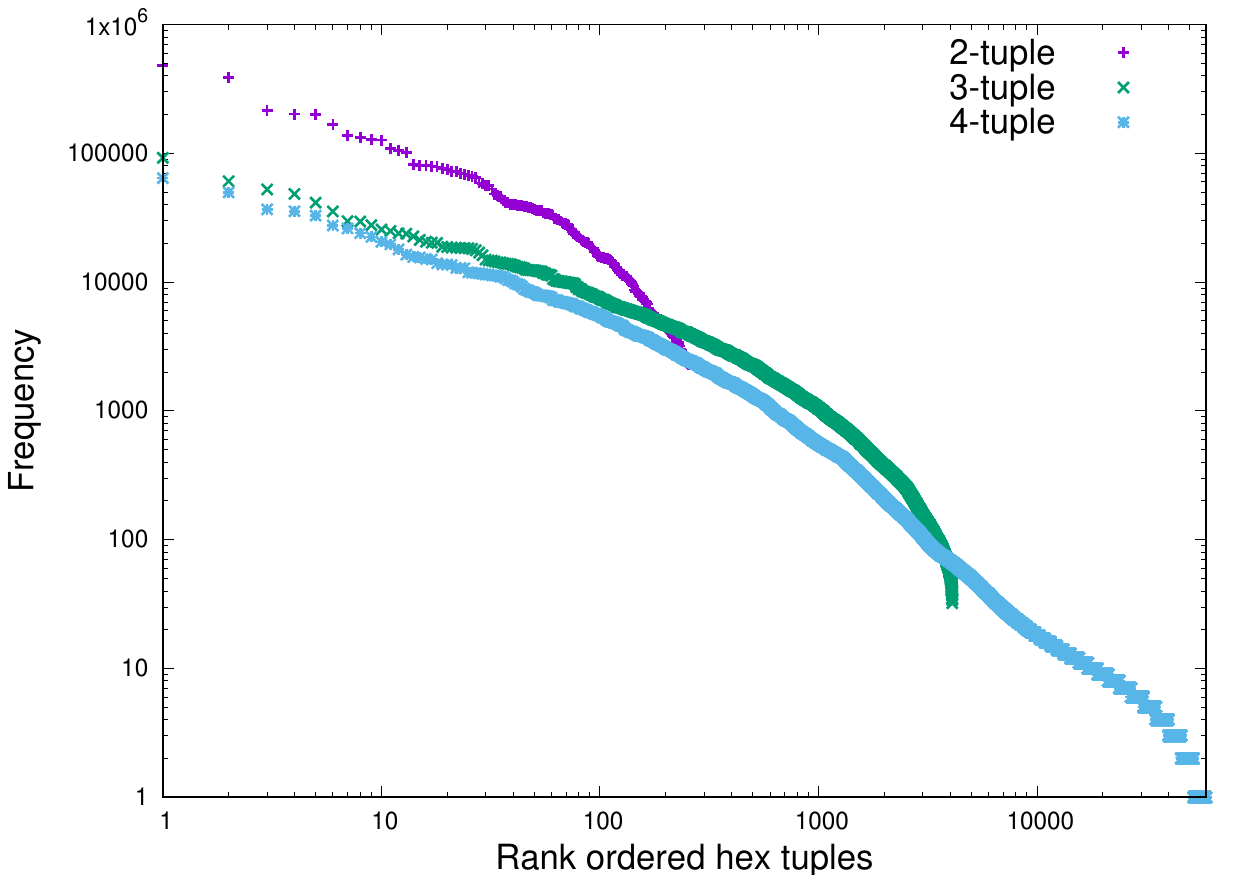}
\caption{The planetarium software \textit{kstars} for 2, 3 and 4-tuples. Data are plotted as the complementary cumulative distribution function.}
\label{fig:combined_kstars_prog}
\end{figure}

Machine-readable computer software is in hexadecimal (base 16) rather than base 4 as in the genome, so a 3-tuple in computer software has as many possibilities as a 6-tuple in the genome.  Fig. \ref{fig:combined_kstars_prog} shows the results of tuple analysis for 2, 3 and 4- tuples using the open source planetarium software \textit{kstars}.  The slopes are steeper than those for any of the genomes shown but are still self-similar and with linearity over 2-3 decades depending on the tuple. These analyses of machine code show the same droop in the tail of the distributions as was seen with the genome distributions and which, as discussed above, is integral to the CoHSI prediction.

\begin{figure}[ht!]
\centering
\includegraphics[width=12cm]{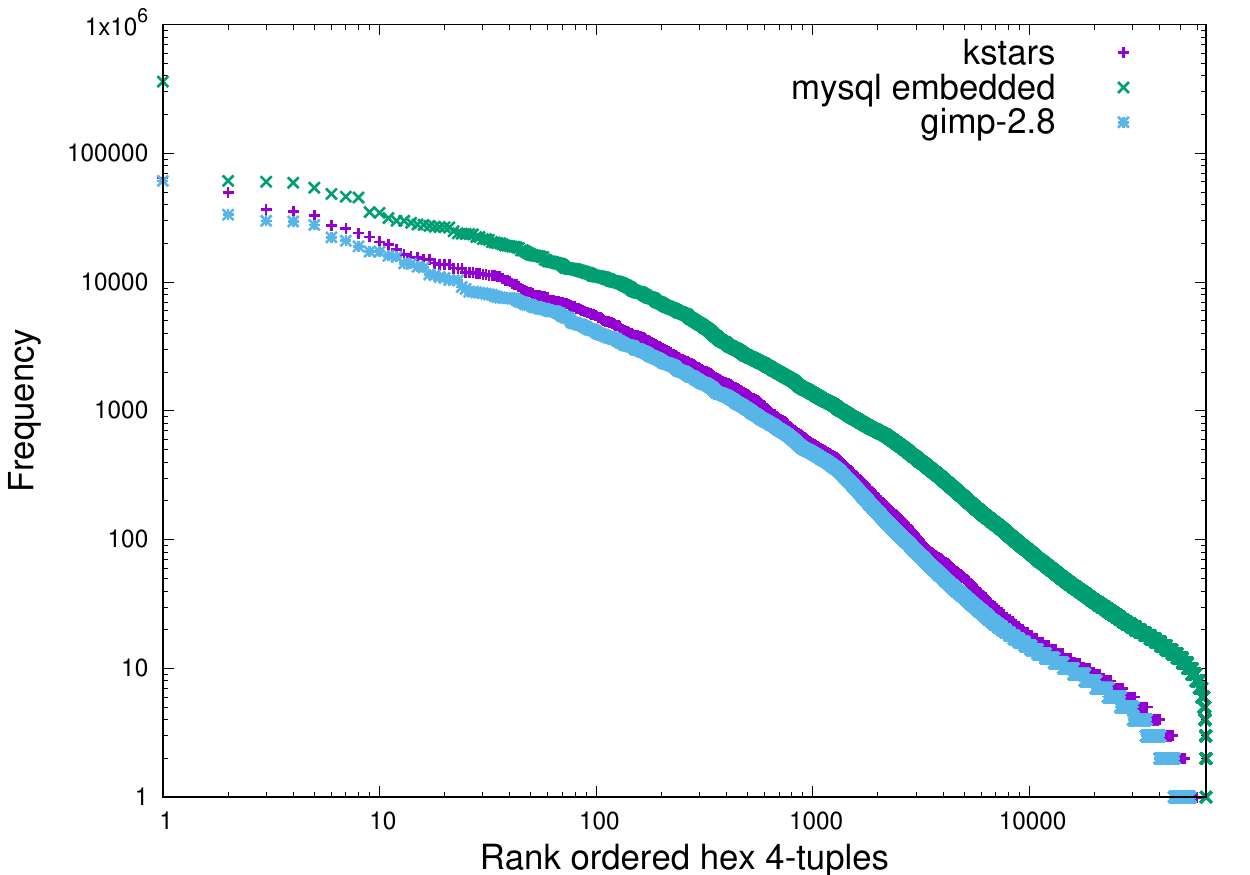}
\caption{The occurrence rates of 4-tuples of machine-readable code in three computer programs, \textit{kstars}, \textit{MySQL} and \textit{gimp}. Data are plotted as the complementary cumulative distribution function.}
\label{fig:combined_4_hex}
\end{figure}

In Fig. \ref{fig:combined_4_hex}, we show 4-tuples for three different programs, the planetarium software \textit{kstars} used above, supplemented with the embedded version of the \textit{MySQL} database software and a version of the well known image editing software, \textit{gimp}.  These programs have a very different function but still exhibit the strongly self-similar near linear behaviour with a droop in the lowest-occupied ranks that we have already consistently seen.  Note that these programs are much smaller than the genomes used earlier giving a little more statistical fluctuation, and that the slope of the power-law is rather steeper. The significance of this is unclear, as in CoHSI models the power-slope is an undetermined Lagrange parameter.

\textit{R lm() reports that the associated p-value matching the power-law linearity for gimp in the 4-tuple ccdf of Fig. \ref{fig:combined_4_hex} is $< (2.2) \times 10^{-16}$ over the 3-decade range $20-20000.0$, with an adjusted R-squared value of $0.975$.  The slope is $-1.36 \pm 0.21$, approximately 3 times steeper than the equivalent slope for the genomic tuples.}

\begin{figure}[ht!]
\centering
\includegraphics[width=12cm]{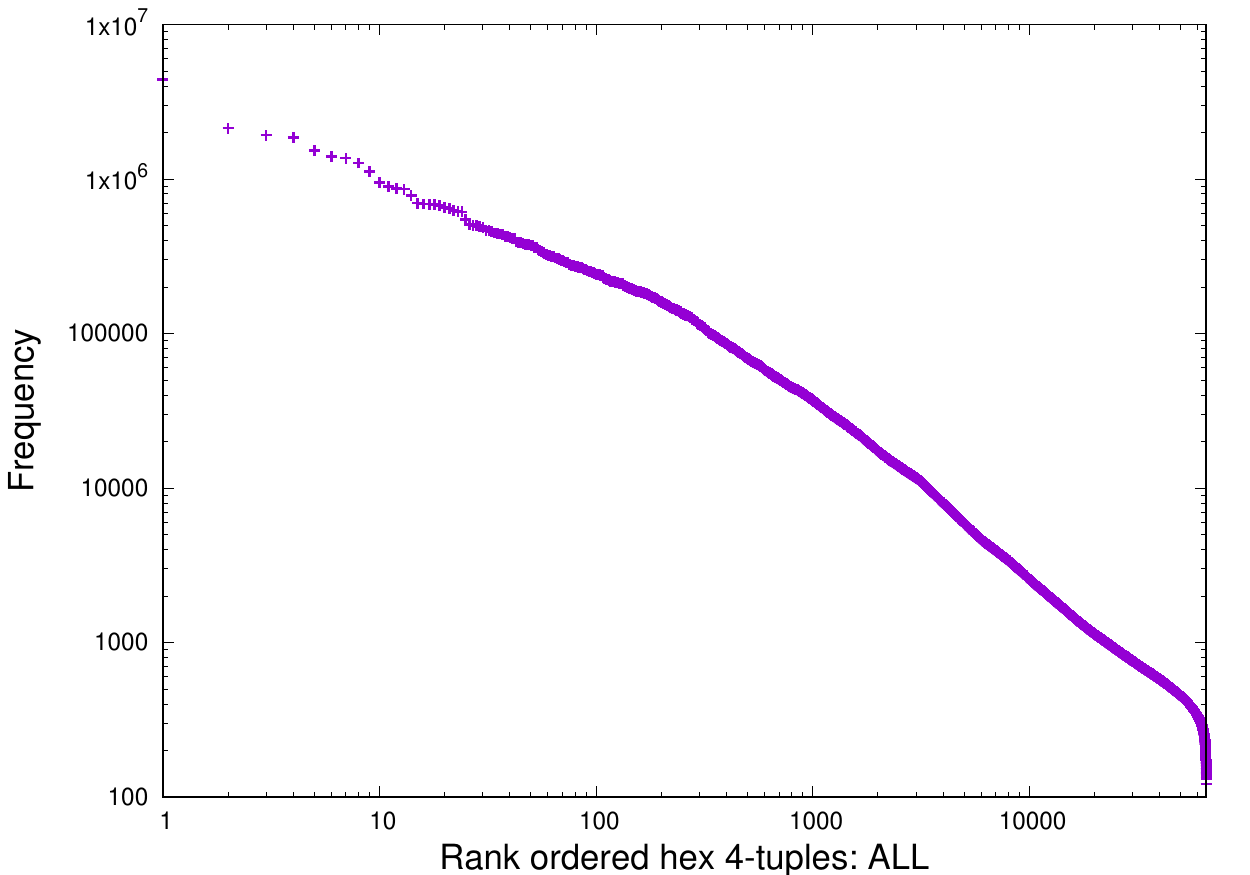}
\caption{The occurrence rates of 4-tuples of machine-readable (hexadecimal) code in 5,307 binary programs located in /usr/bin and /bin of a standard Ubuntu 18.04 LTS distribution treated as a single combined system. Data are plotted as the complementary cumulative distribution function.}
\label{fig:all_4_hex}
\end{figure}

Finally, in Fig. \ref{fig:all_4_hex} when we combine together all 5,307 binary programs in a typical Ubuntu 18.04 LTS distribution totalling some 670MB of binary hexadecimal code combined into a single system, the underlying CoHSI power-law is abundantly visible.

\textit{R lm() reports that the associated p-value matching the power-law linearity in the 4-tuple ccdf of Fig. \ref{fig:all_4_hex} is $< (2.2) \times 10^{-16}$ over the 4-decade range $20-200000.0$, with an adjusted R-squared value of $0.995$.  The slope is $-1.13 \pm 0.08$.}

\subsubsection*{Sensitivity analysis}
A sensitivity analysis of the machine code utilizing offsets of the tuple reading window revealed no significant differences, just as in the genome analysis earlier, and is not therefore shown.

\section*{Conclusions}
The results presented here are an integral part of an ongoing project to test as exhaustively as possible the predictions of a novel conservation principle \cite{HatTSE14,HattonWarr2015,HattonWarr2017}, that has been postulated to constrain the structure of all qualifying discrete systems. This principle (the Conservation of Hartley-Shannon-Information or CoHSI) considers discrete systems constrained not only by their total size, but also by their total information content. By embedding Hartley-Shannon information (in which the symbols are free of meaning) in a statistical mechanical framework, two equations emerge that predict, respectively, the behaviour of the two types of discrete system that have been discovered so far \cite{HattonWarr2017}. The first of these systems is termed \textit{heterogeneous}, and the smallest pieces of the system (tokens, symbols or signs) are assembled sequentially in distinguishable order into larger structures (termed components). Three examples of such heterogeneous systems are proteomes (where amino acids are the tokens and proteins are the components), natural languages (where letters are the tokens and  words are the components) and computer software, in which the programming tokens are assembled into routines or sub-routines, which are the components. The second type of discrete system is termed \textit{homogeneous}, and such systems are characterized simply by the frequency of occurrence of their constituent pieces, as delineated by some consistent categorization, without regard to any distinguishable order. An example of such a system is also provided by natural language texts, when the frequency of each word in a text is counted. Amongst the predictions of CoHSI for heterogeneous systems are scale independence, an overwhelmingly likely (canonical) distribution of component lengths which causes average length to be highly conserved, and the inevitability of very long components, due to a high fidelity power-law tail. The predictions of CoHSI for homogeneous systems reduce simply to a power law relationship with a natural drooping tail for poorly represented categories in the frequencies of the tokens, i.e. Zipf's Law (of which CoHSI provides an alternative proof). 

In previous work \cite{HattonWarr2017} we have shown that proteomes and computer software (as programming languages) both conform closely to the predictions of CoHSI for heterogeneous systems. However, proteomes and software are both higher-level representations of more fundamental symbol-based systems. In the case of proteomes this is the genome (DNA) and for software source code it is the machine-readable code. While we do not wish to pursue the analogy of the genome as computer software, DNA and machine code are both simply long strings of symbols (4 bases in DNA, and the 16 symbols of machine-readable code). Thus both DNA and machine-readable code can both be analyzed as homogeneous discrete systems in terms of the frequency of occurrence of tuples, or short motifs of the symbols of which they are composed.

The predictions of CoHSI for the tuple frequency of DNA and machine-readable code were examined using the sequenced genomes of 10 organisms, ranging from bacteria to human, and the entire binary distribution of an Ubuntu 18.04 LTS distribution. The results were in all cases emphatically as predicted by CoHSI, and for DNA were also independent of offset (reading frame), the proportion of non-protein coding DNA in the genome, or indeed the pooling of the 10 genomes.  For binary computer programs, the results are independent of offset and we made no distinction between control sections of the code and data sections.

Discrete systems and their associated Zipfian properties have been subjected to extensive investigation and several studies have identified similar tuple frequency behaviour to that described here. In this paper, we have restricted our discussion of CoHSI to applications in the genome and in machine code, however we mentioned earlier work by \cite{StephensBialek2010} on text analysis of 4-letter words in the works of Jane Austen and others and this contains interesting and relevant insights.  These authors are particularly interested in semantics and distinguish 4-letter words into several groups a) Those which appear in their text collections, b) Those which are genuine English words but are not used in their text collections and c) Those words which comprise any of the $26^{4}$ ways of arranging the letters of the alphabet for a 4-letter word.  Of these c) dwarves a) and b)  in magnitude but as they show (their Fig. 3), the populations with semantic meaning - a) and a)+b) - obey the power-law with a droop but the amount of droop is inversely proportional to the size of the population of words.  When all patterns are included, the result is power-law with a slight droop as predicted by CoHSI.  Of course, the systems we considered here are considerably larger than the corpus considered by \cite{StephensBialek2010}.

The results of \cite{StephensBialek2010} support our interpretation throughout this and previous papers \cite{HattonWarr2018b} that the CoHSI distribution represents in some sense an equilibrium distribution to which all qualifying discrete systems are drawn and departures from it represent semantic pressure through local mechanism such as natural selection or human volition. 

The power law behaviour of discrete systems  was originally  observed empirically, and Zipf's Law remains controversial with respect to its underlying causes \cite{Piantadosi2014}. In considering the large literature on discrete systems and Zipfian phenomena, it is clear that while many studies have identified the signature of behaviors constrained by CoHSI, the authors have explained the results by an invocation of a wide variety of models \cite{Newman2006,Piantadosi2014}; in contrast, we interpret these studies as supporting the conclusion we advance here for the universal operation of the constraints imposed by CoHSI, even given its extreme parsimony as a theory. In this context it is important to note that \textit{CoHSI is not an assumed model we are trying to fit - it is a predictive theory we are trying to falsify}. 

In discrete systems that conform to Zipf's Law one might predict that a power law relationship should prevail across the full data set from a qualifying system; however, this is rarely if ever observed in natural or experimental data sets and the results that we present here are no exception; we see deviations both at the highest tuple frequencies (typically higher than predicted frequencies) and also with the lowest tuple frequencies, where all of the curves show a pronounced drooping tail. Considering first the highest frequency ranks, two points can be made; first, CoHSI is not a straitjacket, and although it will always guide a system towards an equilibrium state, local departures from the equilibrium can be forced through pressures such as natural selection (in biological systems) semantic pressures (in language) and syntactic and semantic conventions (as in computer programs). If we consider the overpopulation of the highest frequency ranks in the genome, it is noticeable that collectively the most frequent 8-tuple in the eukaryotic species (Table \ref{tab:bigsmall}) is (A)${_8}$ or (T)${_8}$. Poly(A) tracts are known to be associated with the \textit{Alu} repetitive DNA element in primates \cite{Lustig1984} and it is not surprising to find that poly(A) and poly(T) tracts are found in eukaryotic genomes that contain significant numbers of transposable elements that have reverse transcribed intermediates. Transposable elements constitute large portions of eukaryotic genomes, and their continued expansion \cite{Sheinman2016} constitutes, we propose, an evolutionary pressure against which CoHSI is unable to re-establish equilibrium. In a similar way, the over-representation of certain tuples in machine-readable code stems from normal practice in operating system software of loading binary computer programs into memory which has been flash-filled with tracts of 00000000 or ffffffff, many of which remain when the program is dumped for analysis as we have done here. 

Considering now the drooping tails (i.e. the under population of the lowest frequency ranks), although we cannot exclude the influence of stochastic effects we note that the mathematics of the CoHSI equation implies that such a droop will occur in the lowest frequency ranks, and this was confirmed by exact calculation.

The results reported here and in prior studies both by ourselves and others, as discussed above, provide a wealth of data which support the notion that CoHSI constrains the properties of discrete systems in the manner described. Indeed, although CoHSI is not a strait-jacket, we have been unable to falsify any of its predictions thus far, even with systems of such radically different provenance as the genome, and executable machine code.

\section*{Completing the circle}
In this paper, we have examined at the lowest accessible level of categorization two discrete systems of distinct provenance, (bases in the genome and machine-readable hexadecimal characters in computer software) and demonstrated that they have the same CoHSI-predicted global properties.

\begin{figure}[ht!]
\centering
\includegraphics[width=8cm]{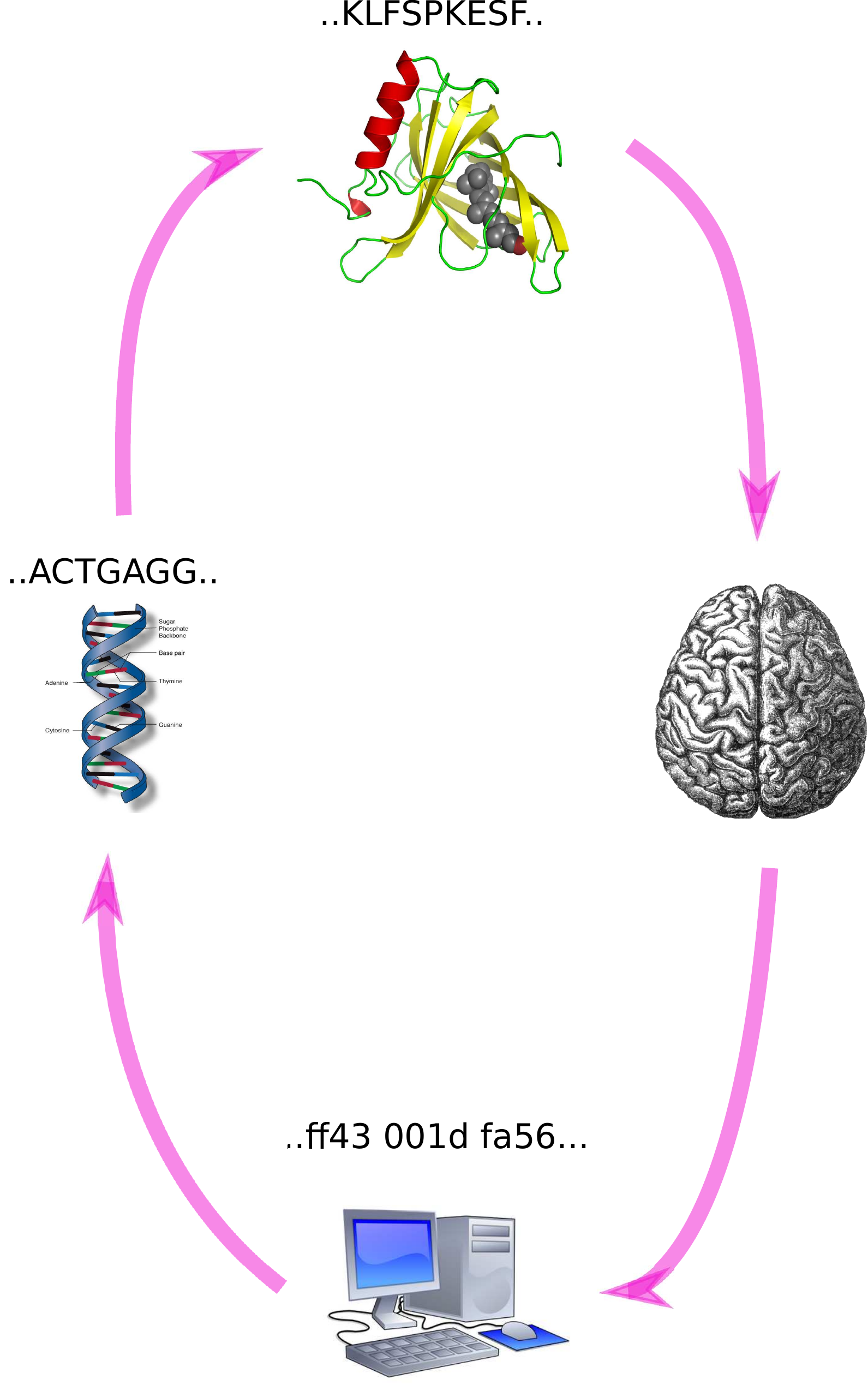}
\caption{Showing the remarkable relationship between 3 separate but intimately linked CoHSI systems; the known set of proteins (\cite{HattonWarr2015}), a heterogeneous system; and the genome and computer software which are homogeneous systems as analyzed here.}
\label{fig:circleoflife}
\end{figure}

Although these systems appear to be of entirely different provenance, arising from the evolution of life (genomes) and from human cognitive processes (computer software) they are inextricably linked at two profound levels. At one level, as illustrated in Fig. \ref{fig:circleoflife} the genome is transcribed into proteins, proteins are one expression of the living systems that include \textit{Homo sapiens}, a species that has developed sufficient intelligence to be able to design silicon-based machines to analyze its own genetic code and to design novel organisms (including humans, \cite{NAP24623}) based on customized and edited DNA sequences, thus completing an iterative cycle that promises an interesting future \cite{Harari2016}. The second level at which genomes and computer software are linked has to do with CoHSI itself. CoHSI is a mechanism-free and token-agnostic ergodic theory, and there is no reason to suppose that the predictions of CoHSI will not constrain all discrete systems, of which the universe is the largest instance  \cite{HattonWarr2017} whereas genomes and machine-code are rather smaller examples. 

\section*{Appendix}
\begin{figure}[ht!]
    \captionsetup[subfigure]{labelformat=empty}
    \centering
    \begin{subfigure}[t]{0.5\textwidth}
        \centering
        \caption{A}
        \includegraphics[width=6cm]{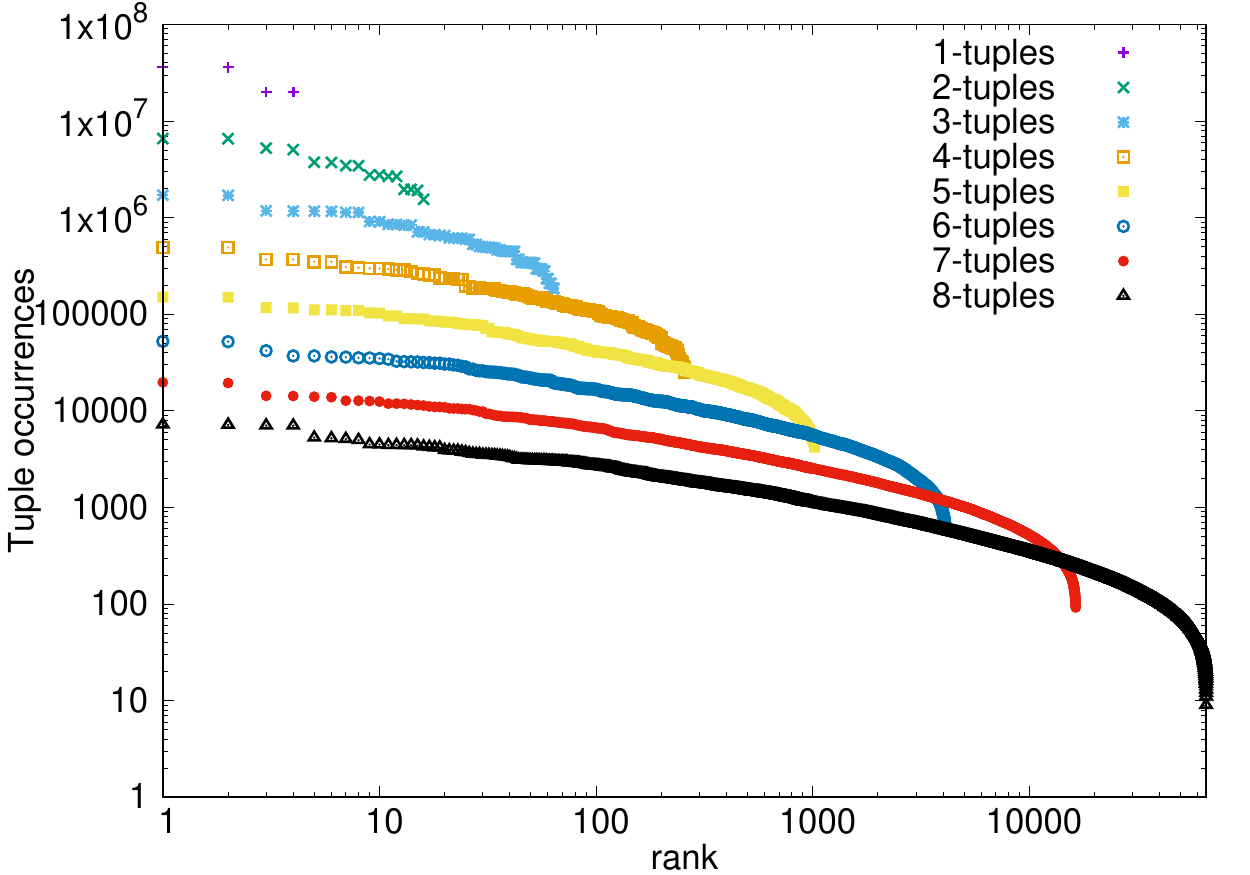}
        \label{fig:ciona}
    \end{subfigure}%
    ~ 
    \begin{subfigure}[t]{0.5\textwidth}
        \centering
        \caption{B}
        \includegraphics[width=6cm]{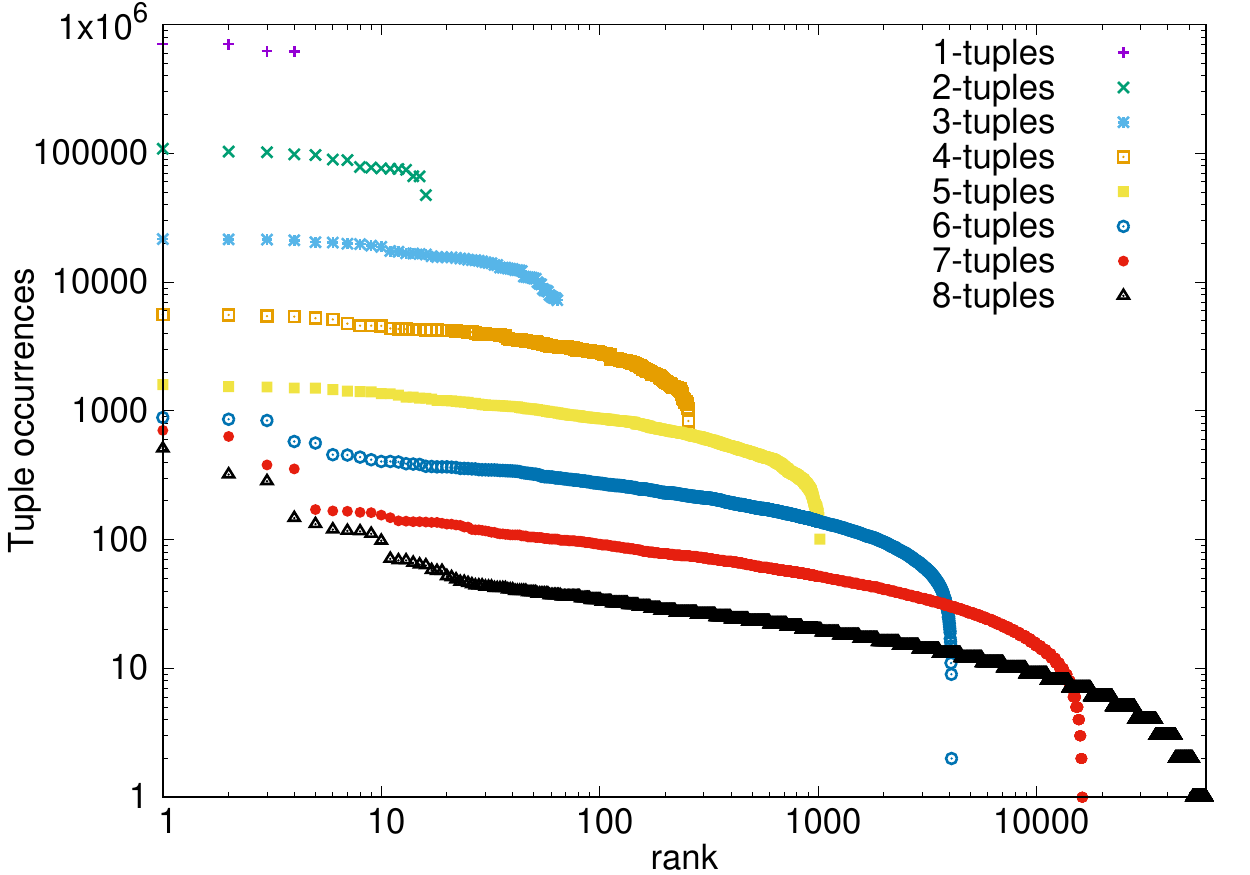}
        \label{fig:cyanobacteria}
    \end{subfigure}%
    
    \captionsetup[subfigure]{labelformat=empty}
    \centering
    \begin{subfigure}[t]{0.5\textwidth}
        \centering
        \caption{C}
        \includegraphics[width=6cm]{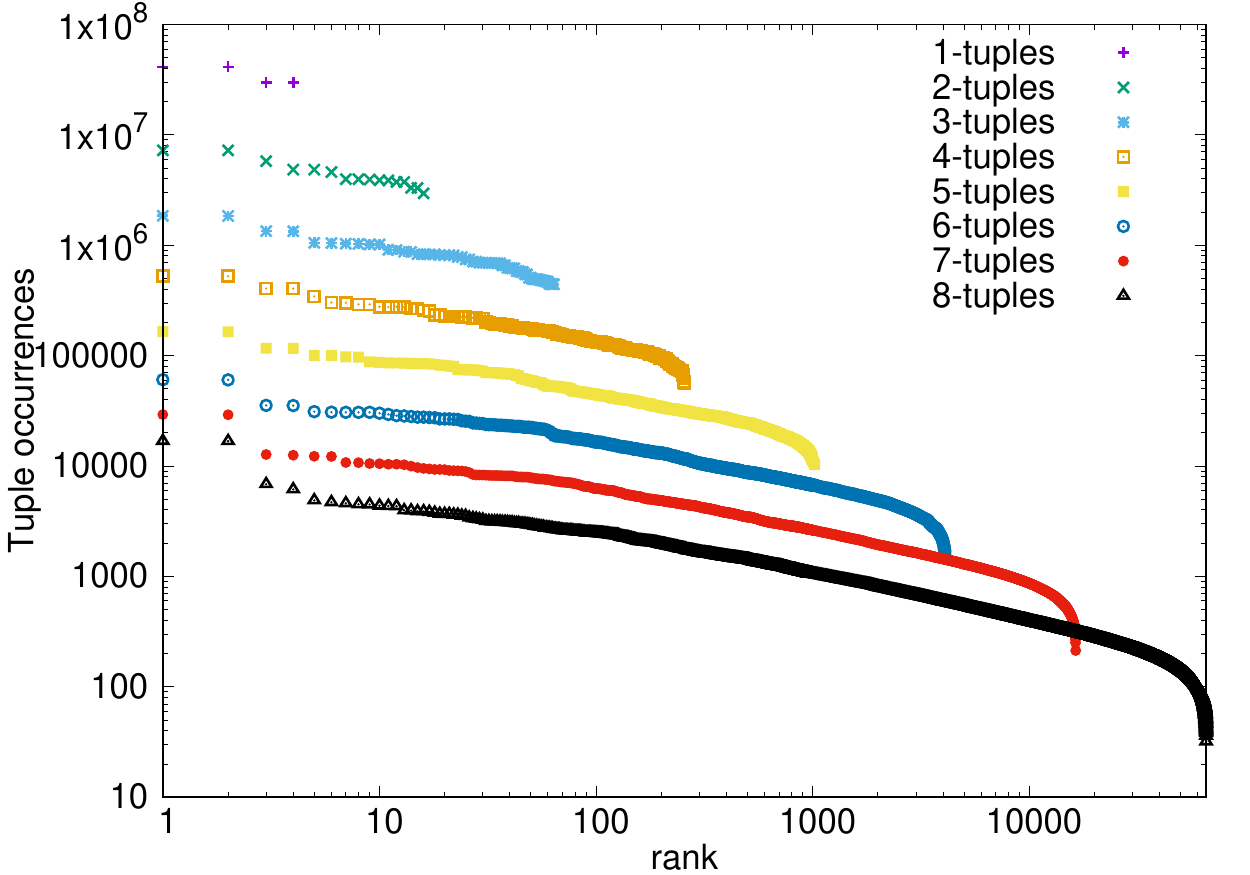}
        \label{fig:fruitfly}
    \end{subfigure}%
    ~ 
    \begin{subfigure}[t]{0.5\textwidth}
        \centering
        \caption{D}
        \includegraphics[width=6cm]{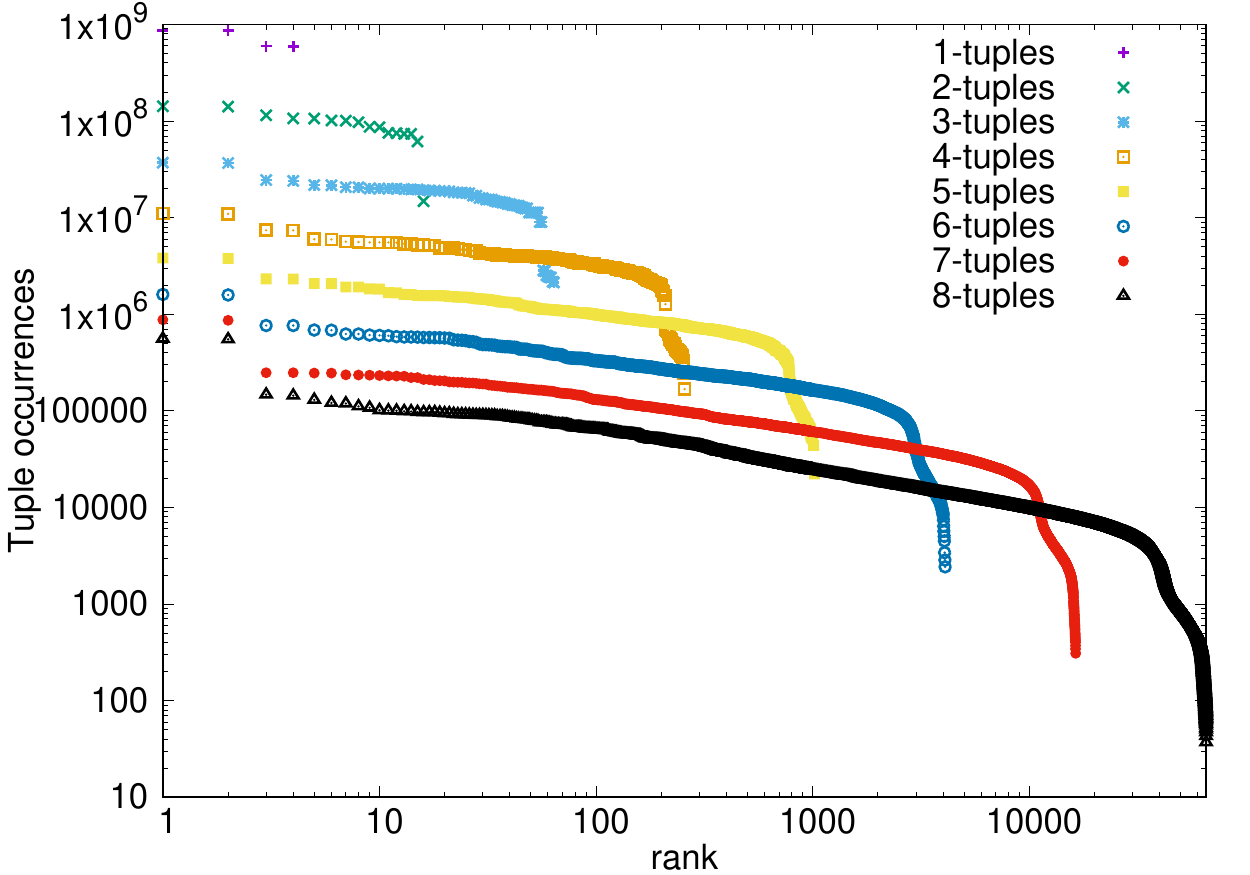}
        \label{fig:gorilla}
    \end{subfigure}%
    
    \captionsetup[subfigure]{labelformat=empty}
    \centering
    \begin{subfigure}[t]{0.5\textwidth}
        \centering
        \caption{E}
        \includegraphics[width=6cm]{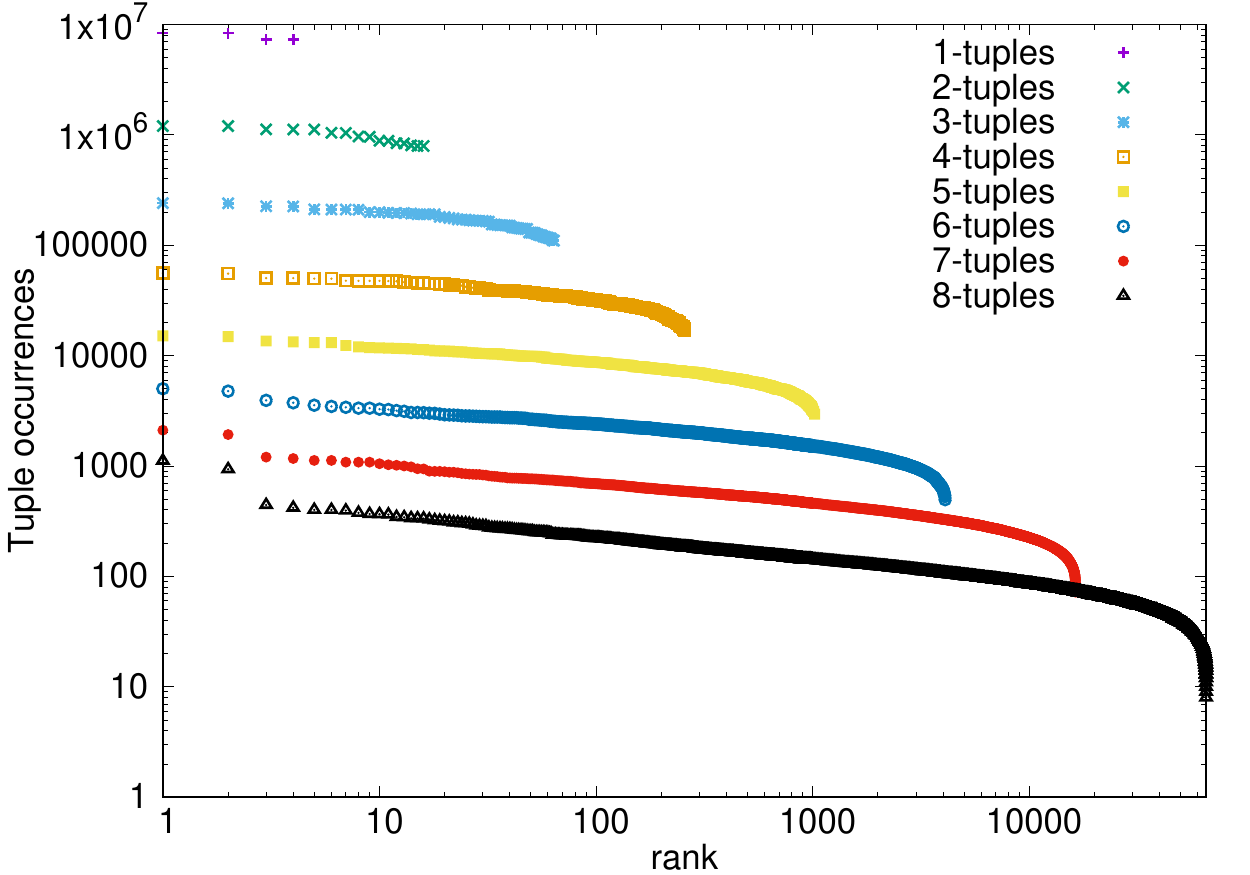}
        \label{fig:mushroom}
    \end{subfigure}%
    ~ 
    \begin{subfigure}[t]{0.5\textwidth}
        \centering
        \caption{F}
        \includegraphics[width=6cm]{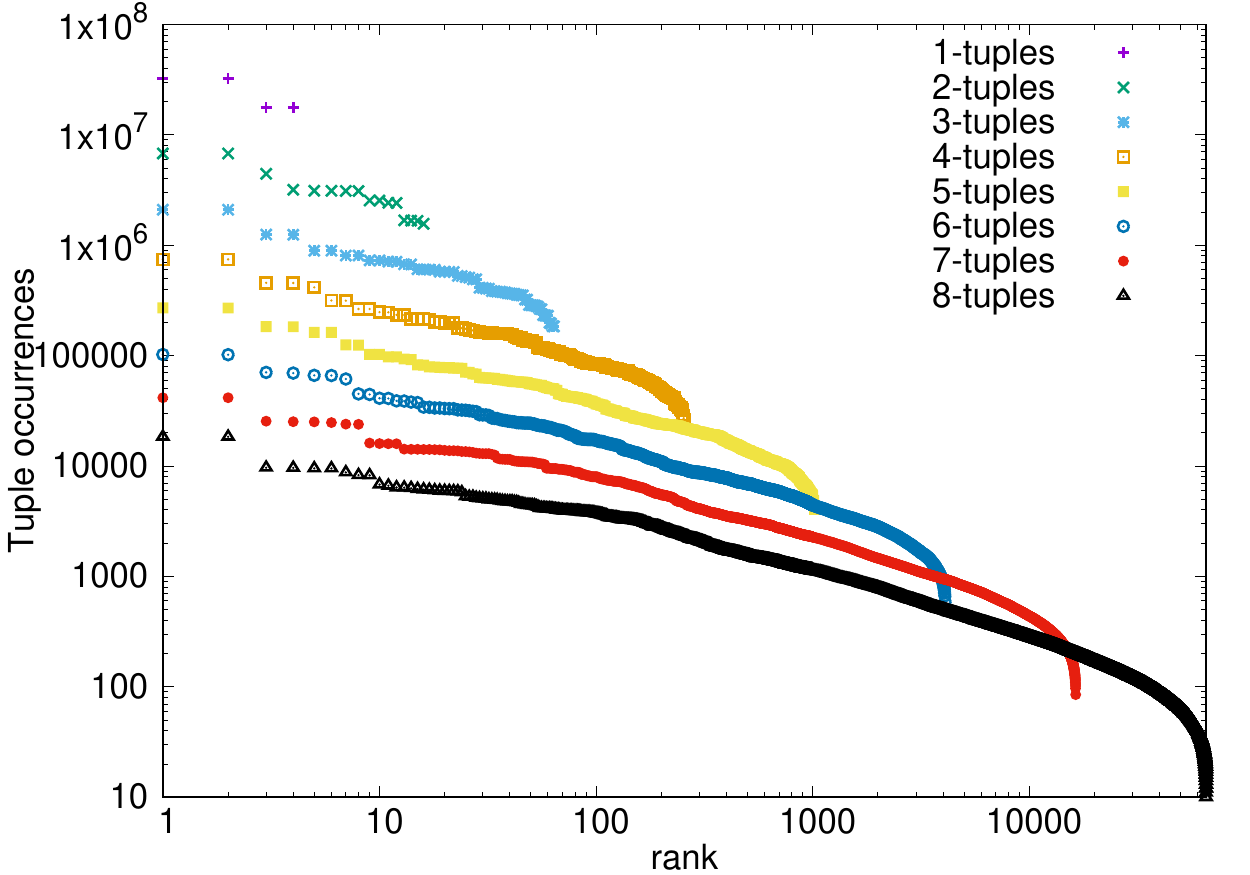}
        \label{fig:nematode}
    \end{subfigure}%
    \caption{n-tuple frequencies for (A) Ciona (B) Cyanobacteria (C) Fruitfly (D) Gorilla (E) Mushroom and (F) Nematode. Data are plotted as the complementary cumulative distribution function.}
\end{figure}

\begin{figure}[ht!]
    \captionsetup[subfigure]{labelformat=empty}
    \centering
    \begin{subfigure}[t]{0.5\textwidth}
        \centering
        \caption{A}
        \includegraphics[width=6cm]{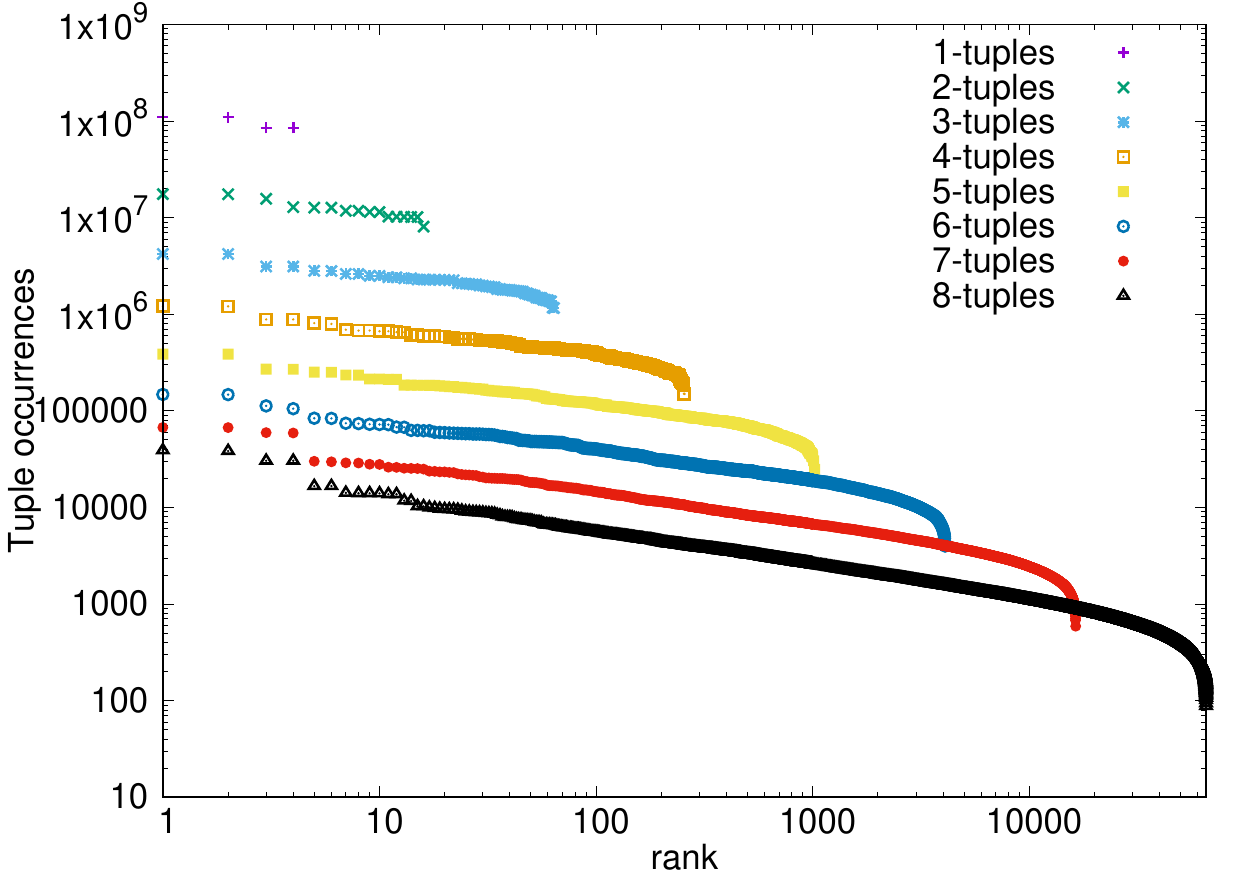}
        \label{fig:rice}
    \end{subfigure}%
    ~ 
    \begin{subfigure}[t]{0.5\textwidth}
        \centering
        \caption{B}
        \includegraphics[width=6cm]{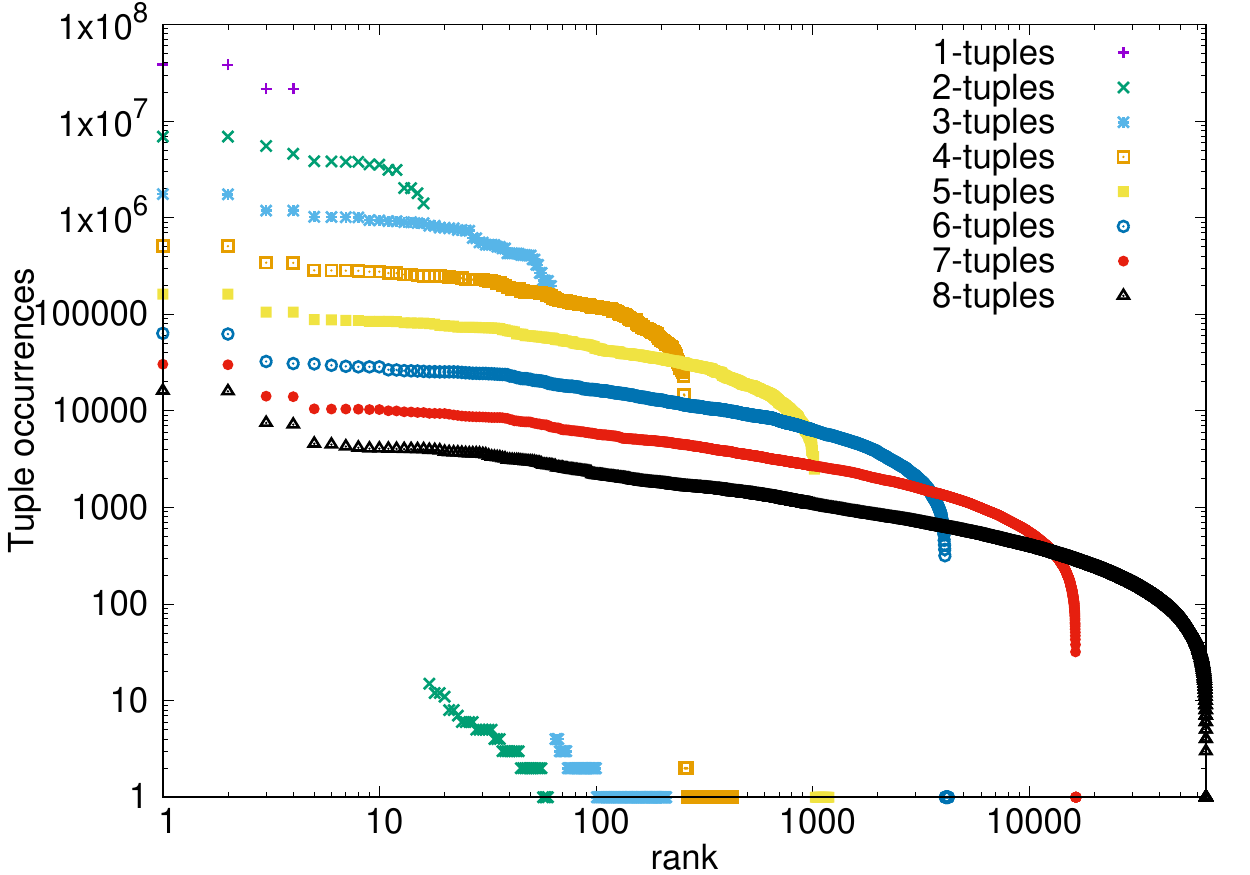}
        \label{fig:thalecress}
    \end{subfigure}%
    
     \begin{subfigure}[t]{0.5\textwidth}
        \centering
        \caption{C}
        \includegraphics[width=6cm]{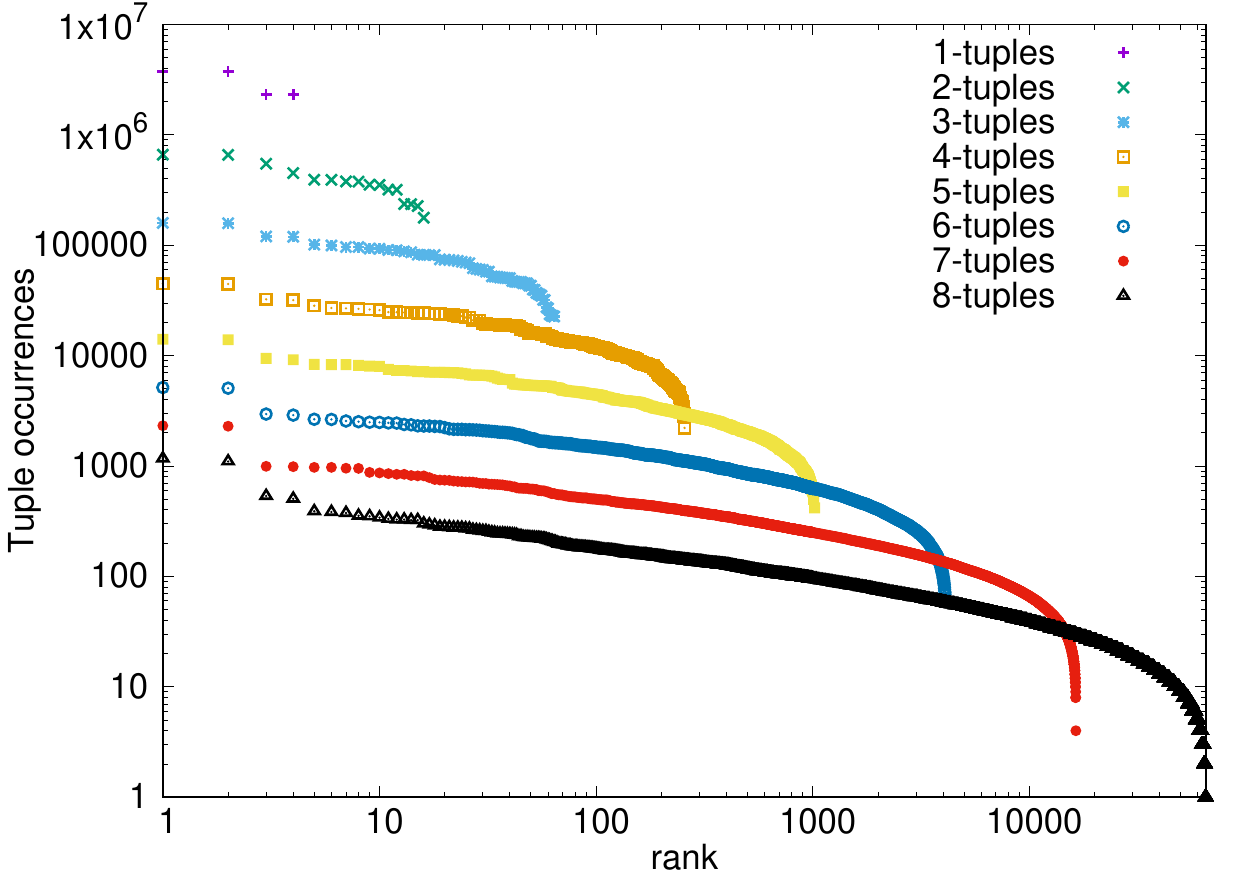}
        \label{fig:yeast}
    \end{subfigure}%

    \caption{n-tuple frequencies for (A) Rice (B) Thale Cress (C) Yeast. Data are plotted as the complementary cumulative distribution function.}
\end{figure}

Here we present the remarkable visual similarity between the 1-8 tuples for the haploid version of each of the remaining genomes analysed in this study, emphatically illustrating the unseen hand of CoHSI in their development.  These are shown alphabetically by species as Figs \ref{fig:ciona} - \ref{fig:nematode} and Figs \ref{fig:rice} - \ref{fig:yeast}.

These figures have many interesting features, quite apart from their extraordinary similarity, for example, the entire Gorilla distribution Fig. \ref{fig:gorilla} is identical to that of the human, Fig. \ref{fig:human}, including the fine structure detail in the tail.  This is also clearly visible when they are compared in Fig. \ref{fig:allslopes}.

\section*{Acknowledgements}
We owe a continuing debt to countless volunteers who provide high quality open source software, (we used Linux and the redoubtable perl),  and also the scientists who enabled open access to the datasets.  Without open source and open data, science would unquestionably descend into a new Dark Age overseen by tribal influences.  In various discussions, Gillian Libretto provided the insights which led to Figure \ref{fig:circleoflife} and we are indebted also to Bob Chapman for numerous penetrating insights and relevant background knowledge of which we were generally ignorant.

Sub-images in Fig. \ref{fig:circleoflife} are displayed under the Creative Commons Licence of Wikipedia.

\begin{itemize}
 \item \textbf{Correspondence:} Correspondence and requests for materials
should be addressed to Les Hatton ~(email: lesh@oakcomp.co.uk).
\end{itemize}

\section*{Author's contributions}
LH performed the analyses, LH and GW developed the arguments, discussed the results and contributed to the text of the manuscript.

\section*{Competing interests}
The authors declare no competing financial interests.

\section*{Funding}
This work was funded by the authors.

\newpage


%
%

\bibliographystyle{alpha}
\bibliography{bibliography}

\end{document}